\documentclass[a4paper,fleqn,usenatbib]{mnras}

\usepackage{newtxtext,newtxmath}

\usepackage[T1]{fontenc}
\usepackage{ae,aecompl}
\usepackage{times}

\usepackage{graphicx}	% Including figure files
\usepackage{amsmath}	% Advanced maths commands

\usepackage{acronym}
\usepackage{ulem}
\usepackage{bm}
\newcommand{\msun}{\mathrm{M}_\odot}
\newcommand{\rsun}{\mathrm{R}_\odot}

\newcommand{\kms}{km~s$^{-1}$}
\newcommand{\pc}{\textrm{per cent}}

\acrodef{sn}[SN]{supernova}
\acrodefplural{sn}[SNe]{supernovae}
\acrodef{ccsn}[ccSN]{core-collapse supernova}
\acrodefplural{ccsn}[ccSNe]{core-collapse supernovae}
\acrodef{ns}[NS]{neutron star}
\acrodef{tzo}[T\.ZO]{Thorne-\.Zytkow object}

\title[NS-companion collision]{Neutron stars colliding with binary companions: formation of hypervelocity stars, pulsar planets, bumpy superluminous supernovae and Thorne-\.Zytkow objects}

\author[R. Hirai et al.]{
Ryosuke Hirai$^{1,2}$\thanks{E-mail: ryosuke.hirai@monash.edu},
Philipp Podsiadlowski$^{3}$
\\
$^{1}$OzGrav: The Australian Research Council Centre of Excellence for Gravitational Wave Discovery, Clayton, VIC 3800, Australia\\
$^{2}$School of Physics and Astronomy, Monash University, VIC 3800, Australia\\
$^{3}$Department of Physics, University of Oxford, Keble Rd, Oxford OX1 3RH, United Kingdom
}

\date{Accepted XXX. Received YYY; in original form ZZZ}

\pubyear{2021}

\begin{document}
\label{firstpage}
\pagerange{\pageref{firstpage}--\pageref{lastpage}}
\maketitle

\begin{abstract}
We perform 3D hydrodynamical simulations of new-born neutron stars (NSs) colliding with main-sequence binary companions after supernova explosions. Based on those hydrodynamical models, we construct a semi-analytical formula that describes the drag force inside stars with steep density gradients. We then compute the outcome of NS--companion collisions over a wide range of parameters using the semi-analytical formula. Depending on the direction and magnitude of the natal kick, we find that the collision may lead to various outcomes. For relatively fast kicks and high impact parameters, the NS may penetrate the companion star envelope without merging. By allowing the NS to plunge through companions, the companion can be accelerated to have runaway velocities up to $\sim10~\pc$ above the theoretical upper limit considered in classical binary disruption scenarios. The NS can capture and carry away up to a few $\pc$ of the companion envelope as it escapes, which may form pulsar planets or cause outflows through accretion to heat the ejecta from inside and power the supernova light curve. For lower impact parameters, the neutron star will directly merge with the companion and form a Thorne-\.Zytkow object. In intermediate cases, the NS penetrates the companion envelope several times before merging, possibly causing multiple bumps in the supernova light curve like in SN2015bn and SN2019stc.
\end{abstract}


%%%%%%%%%%%%%%%%%%%%%%%%%%%%%%%%%%%%%%%%%%%%%%%%%%

%%%%%%%%%%%%%%%%% BODY OF PAPER %%%%%%%%%%%%%%%%%%

\section{Introduction}\label{sec:introduction}
\Acp{ns} are the compact remnants of massive stars that die with an energetic \ac{ccsn} explosion. It is known observationally that they are usually given a strong natal kick at birth \citep[]{lyn94}, induced by the asymmetric ejection of matter or neutrinos \citep[see][and references therein]{lai01}. Typical velocities of natal kicks are a few 100~\kms \citep[]{lyn94,hob05,bra16,ver17,kat18,igo21}, while some \acp{ns} can reach velocities in excess of $\gtrsim1,000$~\kms. For example, the projected kick velocity of the central compact object in the \ac{sn} remnant Puppis A is $\sim760$~\kms, implying an even higher physical velocity \citep[]{may20}.

The majority of massive stars are known to be born in binary or multiple systems \citep[]{san12,san14}. When one of the stars in the binary explodes, the orbital parameters can be significantly altered due to the sudden mass loss and the natal kick. Depending on the magnitude and direction of the kick, the binary can survive with wider or tighter orbits, or be disrupted to become two separate stars. If it survives, the binary can sometimes be observed as an X-ray binary, depending on whether the post-\ac{sn} orbit is tight enough to cause Roche lobe overflow \citep[]{tau06} or sufficient wind accretion \citep[]{RH21b}. A small fraction of those systems may experience a second \ac{sn} and go on to form double \ac{ns} systems that can be observed as binary pulsars \citep[]{hul75} or gravitational wave sources \citep[]{abb17}.

In cases where the binary is disrupted by the first \ac{sn}, the companion star can be hurled out and be observed as runaway stars \citep[``Blaauw mechanism'';][]{bla61}. Most runaway stars are expected to have relatively low velocities \citep[$\lesssim30$~\kms;][]{ren19}, but observations show that some stars are moving with proper motions exceeding $\gtrsim400$~\kms \citep[]{bro05,bro14,irr18b,irr18a}. While at least some of these ``hypervelocity'' stars are consistent with originating from the Galactic centre accelerated via the so-called ``Hills mechanism'' \citep[]{hil88}, more than a handful of them firmly originate from the Galactic disk, which means that they were likely ejected via binary disruption \citep[]{irr10,irr18a,irr19}. However, some of the disk hypervelocity stars reach ejection velocities exceeding the theoretical limit for the Blaauw mechanism \citep[$\sim540$~\kms;][]{tau15}, posing a serious problem on how these hypervelocity stars were formed.

In some rare situations, the \ac{ns} kick may be directed in a way such that it skims the surface of the binary companion or even collide with it. It is typically expected that as the NS collides with the envelope, it will rapidly dissipate its orbital energy and spiral in towards the centre. This has been proposed to be one of the ways to form \acp{tzo}, which are hypothetical objects of \acp{ns} embedded in diffuse hydrogen-rich envelopes \citep[]{tho75,tho77}.

It is not trivial whether the \ac{ns} will immediately fall to the centre, even if it collides with the star. The mass distribution in main-sequence stars are quite centrally concentrated, especially when the envelope is radiative. When \acp{ns} plunge into such envelopes, the drag force may not be strong enough to sufficiently reduce the relative velocity and instead it may penetrate through. By considering envelope penetration, it not only reduces the probability of forming \acp{tzo}, but it can also relax the upper limit for the runaway velocities achieved by the Blaauw mechanism. Whether this is possible depends on how much the \ac{ns} is decelerated by the stellar material. In \citet{RH21} (Appendix C), we explored a similar situation where a compact helium star plunges through a massive main-sequence star. We applied a simple drag formula that ignores the effects of the steep density gradient and stellar response to estimate the amount of deceleration. Our simple models showed that the envelope can be easily penetrated, but it is not clear how appropriate our simplifications are. 3D hydrodynamical simulations are required to evaluate these effects.

There are several other byproducts that could occur in envelope penetrating cases. As the \ac{ns} penetrates the envelope, it will carry away part of the envelope matter with some angular momentum. Part of the captured material can accrete onto the NS by transporting out the angular momentum through an accretion disk. The \ac{sn} ejecta are expected to be still dense and optically thick, so any X-ray emission or kinetic outflows could be reprocessed to power the \ac{sn} light curve. The outer parts of the disk may then go on to form protoplanetary disks, eventually creating pulsar planets \citep[]{wol92}. 

While there are detailed studies of collisions of main-sequence stars with white dwarfs \citep[]{ruf90,ruf92,ruf93} and recently with black holes \citep[]{kre22,ros22}, there is surprisingly little literature on collisions with \acp{ns}. In this paper, we investigate the various outcomes of \acp{ns} colliding with binary companions after \ac{sn}, with particular emphasis on envelope penetrations. We perform 3D hydrodynamical simulations to see how the NS interacts with the companion star. This paper is structured as follows. We first outline the method and setup of our 3D hydrodynamical simulations in Section~\ref{sec:methods}. Simulation results are presented in Section~\ref{sec:results}. We then discuss extensions to the broader parameter space in Section~\ref{sec:discussion} by constructing a semi-analytic model that accurately describes the simulation results. We speculate on applications to various observations in Section~\ref{sec:application}. Summary and conclusions are provided in Section~\ref{sec:conclusion}.

\section{Methods}\label{sec:methods}
\subsection{3D hydrodynamical simulation}

We carry out 3D hydrodynamical simulations using the hydrodynamics code \texttt{HORMONE} \citep[]{RH16}. It is an Eulerian hydrodynamics code that solves the Euler equations based on a Godunov scheme with the HLLC approximate Riemann solver for the numerical flux. Self-gravity is treated by solving the Poisson equation for the gravitational field as the initial condition and then evolving the field with a wave equation \citep[hyperbolic self-gravity solver;][]{RH16}. It is known to have negligible errors as long as the gravity propagation speed is set to a value larger than the maximum characteristic speed in the computational domain. We assume an ideal gas equation of state with an adiabatic index $\gamma=5/3$ for all of our simulations.

For the initial condition, we place a stellar model at the origin of a spherical coordinate grid. The computational domain in \texttt{HORMONE} is covered with a non-uniform grid with 600 grid points in the radial direction, in which the grid spacing is set so that it is at least 5 times smaller than the local pressure scale height in the initial star ($\Delta r\leq H_p/5$). We impose a limit on the variation between neighbouring grid sizes to be less than 2 per cent ($|\Delta r_n/\Delta r_{n+1}-1|<0.02$). Outside the star, the spacing is increased as a geometrical series. As a result, the grid size is smallest around the surface of the star where the scale height is shortest and increases both inwards and outwards. The outer boundary is set at $\sim$40 times the stellar radius and an outgoing boundary condition is applied. For the angular grid, we divide the polar direction uniformly in $\cos\theta$ and uniformly in the azimuthal direction. This ensures that the solid angle of every cell is equal, and has higher resolution around the equatorial plane. We assume equatorial symmetry to further reduce the computational cost by a factor 2. To avoid severe Courant conditions, we effectively reduce the angular resolution around the coordinate origin. We first set the innermost 2 radial cells to be spherical. The following 2 radial cells are divided with an effective angular resolution of $(N_\theta,N_\phi)=(2,8)$, and then we sequentially double the effective angular resolution every 3 radial cells outwards. The highest resolution is $(N_\theta,N_\phi)=(64,256)$. This nested grid structure is treated by computing the hydrodynamics with the highest resolution grid and then averaging conserved quantities over the cells within the effective cells after every time step. Vector quantities are averaged so that the linear momentum is conserved. With this grid, we confirmed that the star is stable and does not expand or contract as a single star for at least $50$~ks, much longer than the duration of all our simulations.

Assuming that the primary star exploded as a \ac{sn} and immediately created a \ac{ns} with some natal kick, we place a point mass at a distance $a$ from the centre of the grid. For a given kick direction ($\theta_\mathrm{kick}, \phi_\mathrm{kick}$) and magnitude $v_\mathrm{kick}$, we calculate the velocity vector in the frame of the companion star and apply that to the \ac{ns} in addition to the pre-\ac{sn} orbital velocity. We then rotate the frame so that the \ac{ns} velocity points in the equatorial plane. This justifies our assumption of equatorial symmetry, as long as the star is spherically symmetric and non-rotating.

The new-born \ac{ns} is treated as a point mass with a softened gravitational potential. We use a cubic-spline kernel for the softened potential in the form introduced in \citet{pri07}. This gravity interacts with the gas only gravitationally, treated as an external force in the Euler equations. We adaptively adjust the softening length so that it is always resolved by $\gtrsim3$ grid points, to ensure energy and momentum conservation. This generally leads to softening lengths much larger than the physical size of the \ac{ns}, and therefore may underestimate the feedback from the accretion onto the \ac{ns}. However, the true effect of accretion feedback is highly uncertain anyway, so we believe this methodology is sufficient for our current purpose. With this procedure, the total energy, linear momentum and angular momentum are all conserved within $\lesssim1~\pc$ for our $M_2=1,~5~\msun$ models. The angular momentum conservation in our $M_2=10~\msun$ model is worse ($\lesssim5~\pc$), but the energy conservation is still within $\lesssim1~\pc$.

In our methodology, we ignore two effects which may have non-negligible impact on our results: stellar rotation and ejecta-companion interaction. If the pre-\ac{sn} stellar rotation is significant, it would be distorted from the spherical shape we assume here. Also, the interaction with the \ac{ns} could be quite different, because the relative velocity between the \ac{ns} and stellar envelope material will be changed. The impact of \ac{sn} ejecta on the companion has been suggested to inflate the star significantly due to heat injection \citep[]{RH15,RH18,oga21}. However, the inflation only occurs for the surface material, so the bulk of the star remains relatively unaffected. Taking these effects into account will unnecessarily increase the number of model parameters. In this paper, we focus only on variations depending on the companion star mass and \ac{ns} kick, and leave any further parameter studies for future work.

\subsection{Stellar model and kick parameters}
We generate the companion star models using the stellar evolution code MESA \citep[]{MESA1,MESA2,MESA3,MESA4,MESA5}. For simplicity, we use a zero-age main-sequence star model with a metallicity $Z=0.014$. In reality, the companion should have a slightly different structure from the zero-age as it has an age equivalent to the lifetime of the primary star. It also may have experienced mass accretion from the primary star in the prior evolution. Both of these effects make only moderate differences to the stellar structure, and shall not significantly influence our results.

The primary star (\ac{sn} progenitor) is fixed to $M_1=6~\msun$, representing a stripped-envelope \ac{sn} progenitor. We assume that the \ac{sn} always produces a $M_\mathrm{NS}=1.4~\msun$ \ac{ns}, meaning that the ejecta mass is roughly $M_\mathrm{ej}\sim4.5~\msun$\footnote{There is a $\mathcal{O}(0.1~\msun)$ difference between the baryonic and gravitational mass of a NS.}, which is consistent with the typical observed ejecta mass for stripped-envelope \acp{sn} \citep[]{can13,tad15,lym16,tad18}. We choose 3 different companion masses: $M_2=1, 5, 10~\msun$. For each companion model, we assume it was orbiting the pre-\ac{sn} primary star on a circular orbit. We choose an appropriate orbital separation $a$ which is as small as possible without allowing the companion star to overflow its own Roche lobe. We then select several kick angles $(\theta,\phi)$ per binary model. The binary and kick parameters for our 3D simulations are summarised in Table~\ref{tab:parameters}. Kick angles are defined in the same way as in \citet{tau98}: the polar axis ($\theta=0$) is pointed in the direction of the orbital motion of the primary and $\phi=0$ sweeps the orbital plane in the direction of the secondary. We fix the kick magnitude to $v_\mathrm{kick}=1,000$~\kms. This is close to the upper edge but within the range of observationally inferred\footnote{It should be noted that there is an observational bias against faster kick velocities, due to the shorter dwell time in the Galaxy.} \citep[]{hob05,ver17,igo21} and theoretically achievable kick velocities \citep[]{jan17}. We also list the periastron distance (normalised by the stellar radius) and eccentricity of the post-\ac{sn} orbit. The periastron distance is calculated assuming a two-body problem of two point particles without any drag. 

\begin{table}[h]
 \begin{center}
  \caption{Model parameters for the 3D hydrodynamical simulations.\label{tab:parameters}}
  \begin{tabular}{cccccccc}
   \hline
   Model & $M_2$ & $R_2$ & $a$ & $\theta$ & $\phi$ & $a_\mathrm{per}$ & $e$\\
         & $(\msun)$&$(\rsun)$&$(\rsun)$&(rad)&(rad)&$(R_2)$& \\\hline
   M1-1  & 1  & 0.94 &  5 & 2.400 & 0.1 & 1.19 & 1.59 \\
   M1-2  & 1  & 0.94 &  5 & 2.250 & 0.1 & 0.46 & 1.32 \\
   M1-3  & 1  & 0.94 &  5 & 1.875 & 0.1 & 0.98 & 2.32 \\
   M5-0  & 5  & 2.52 &  8 & 2.550 & 0.1 & 1.01 & 1.14 \\
   M5-1  & 5  & 2.52 &  8 & 2.400 & 0.1 & 0.53 & 1.17 \\
   M5-2  & 5  & 2.52 &  8 & 2.250 & 0.1 & 0.19 & 1.10 \\
   M5-3  & 5  & 2.52 &  8 & 1.875 & 0.1 & 0.45 & 1.54 \\
   M5-4  & 5  & 2.52 &  8 & 2.475 & 0.1 & 0.76 & 1.17 \\
   M10-1 & 10 & 4.82 & 15 & 2.400 & 0.1 & 0.78 & 1.39 \\
   M10-2 & 10 & 4.82 & 15 & 2.250 & 0.1 & 0.36 & 1.25 \\
   M10-3 & 10 & 4.82 & 15 & 1.875 & 0.1 & 0.29 & 1.39 \\
   M10-4 & 10 & 4.82 & 15 & 2.100 & 0.1 & 0.11 & 1.10 \\
   \hline
  \end{tabular}
 \end{center}
\end{table}

\subsection{Definition of bound and ejected mass}
For the analysis of our results, we define the remaining star, ejected mass and accreted mass in the following way. At each time step, we calculate the centre of mass of the gas (excluding the NS) to define the centre of the star. Then for each cell, we mark that it is bound to the star if $\epsilon+|\bm{v}-\bm{v}_\mathrm{com}|^2/2+\phi_\mathrm{gas}<0$, where $\epsilon$ is the specific internal energy, $\bm{v}, \bm{v}_\mathrm{com}$ are the velocity of the cell and gas centre of mass respectively, and $\phi_\mathrm{gas}$ is the gravitational potential of the gas. Similarly, we mark a cell as bound to the NS when $\epsilon+|\bm{v}-\bm{v}_\mathrm{NS}|^2/2+\phi_\mathrm{NS}<0$, where $\bm{v}_\mathrm{NS}$ and $\phi_\mathrm{NS}$ are the velocity and gravitational potential of the NS respectively. When a cell is marked bound to both the NS and the star, we choose to associate it with the closer component. After flagging every cell to being bound to the NS, star or neither, we integrate the mass over each category to obtain the captured mass by the NS, remaining stellar mass and ejecta mass respectively.

\section{Results}\label{sec:results}
\begin{figure*}
 \centering
 \includegraphics[width=\linewidth]{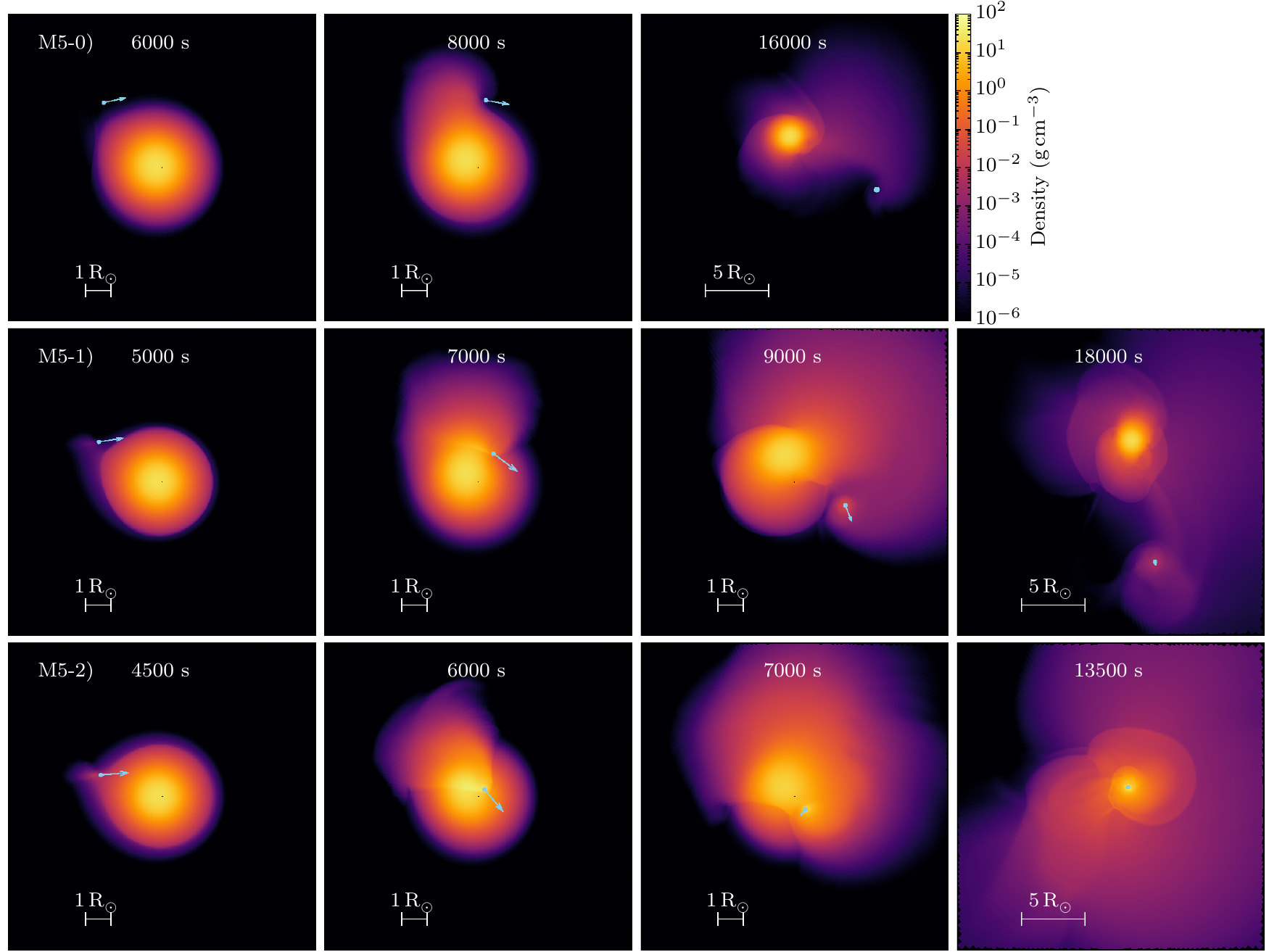}
 \caption{Equatorial density snapshots of the 3D hydrodynamical simulations with a $M_2=5~\msun$ companion and three different kick directions (models M5-0, M5-1, M5-2). Light blue circles indicate the position of the \ac{ns} and the length of the arrows are proportional to the velocity in the simulation frame. Time labels indicate the time since \ac{sn}. The \acp{ns} shoot in from the left of each panel. The top row (M5-0) is a case where the \ac{ns} nominally does not touch the companion envelope. It excites a tidal bulge that later falls and shocks the stellar surface. The middle row (M5-1) shows an example of envelope penetration. There is a clear cloud of material that is carried away from the star by the \ac{ns}. The bottom row (M5-2) corresponds to an immediate merger, where the \ac{ns} never resurfaces from the companion. The end product is a \ac{tzo}.\label{fig:density_5Msun}}
\end{figure*}

We display some snapshots of the hydrodynamical simulations with $M_2=5~\msun$ companions in Figure~\ref{fig:density_5Msun}. 

The top panels are for model M5-0, where the periastron distance is larger than the companion radius. Although it does not touch the star, the gravity of the \ac{ns} creates a tidal bulge that then falls back onto the star, driving shocks. \citet{mac22} described this as ``tidal wave-breaking'', which could be an efficient mechanism in heating and torquing the stellar atmosphere. In eccentric binaries, the repeated tidal heating could lead to enhanced mass loss. In our case, the \ac{ns} is on an unbound orbit, so the heating and torquing effect is minimal.

The bottom row in Figure~\ref{fig:density_5Msun} show results for model M5-2 where the periastron distance is relatively small. A bow shock is formed ahead of the \ac{ns} as soon as it touches the companion star envelope. The shape  of the bow shock is slightly distorted from being axisymmetric due to the density gradient within the star. The gravitational drag rapidly decelerates the NS and it quickly falls to the centre. The dissipated orbital energy is transferred to the envelope through shocks, making it rapidly expand. Such objects are expected to turn into \acp{tzo}.

The middle row shows model M5-1, which is an intermediate case between M5-0 and M5-2 in terms of periapsis. Similar to M5-2, the NS creates a bow shock as it interacts with the envelope. The shock is more distorted with respect to the orbital motion, as the density gradient is steeper in the outer parts of the star. In this model the gravitational drag is not strong enough to capture the NS, so it plunges out of the envelope. The outer parts of the star are heated and torqued by the bow shock and self-interaction shocks from the tidal wave-breaking. Part of the envelope material is captured in the gravitational potential of the NS, forming a torus supported by pressure and centrifugal force.

\begin{figure*}
 \centering
 \includegraphics[width=\linewidth]{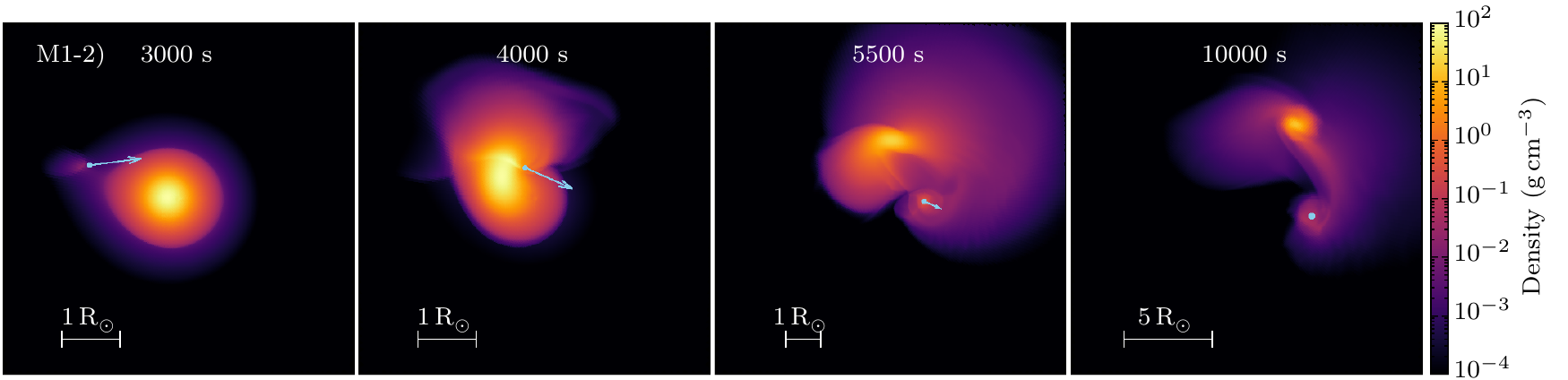}
 \caption{Equatorial density snapshots of the simulation with a $1~\msun$ companion (model M1-2). This resembles a partial tidal disruption event.\label{fig:density_1Msun}}
\end{figure*}

For all other models with the $M_2=5~\msun$ and $10~\msun$ companion, the overall dynamics can be classified into one of the above: (1) tidal distortion + wave-breaking, (2) envelope penetration, and (3) immediate merger. The behaviour of the models with $M_2=1~\msun$ are also qualitatively similar. We show snapshots of the M1-2 model in Figure~\ref{fig:density_1Msun}. The NS penetrates the companion in a similar way to M5-1, but the star is much more tidally distorted straight after the interaction (third panel). Because the mass of the companion is smaller than the NS, this can be interpreted as a weak form of a partial tidal disruption event \citep[]{gui13,ryu20,kre22}. We find no full tidal disruptions within our models due to the relatively low mass ratio ($M_\mathrm{NS}/M_2\lesssim1.4$).

\begin{figure}
 \centering
 \includegraphics[width=\linewidth]{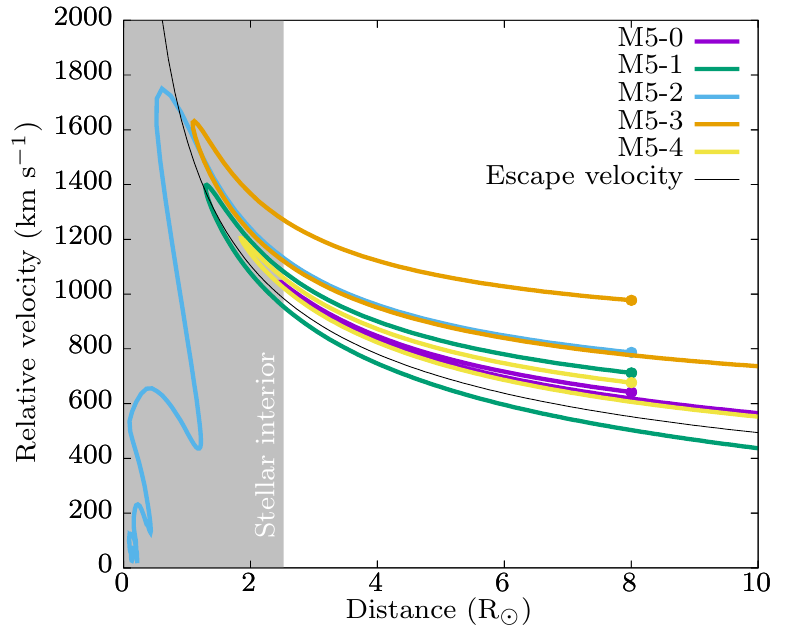}
 \caption{Relative velocity as a function of distance between the NS and companion star for the $M_2=5~\msun$ models. The starting points are marked with circles. The grey shaded region indicates the stellar interior.\label{fig:relative_velo_5Msun}}
\end{figure}

We find that the envelope penetration cases can be further classified into two sub-categories depending on whether the post-penetration orbit is bound or unbound. This can be seen in Figure~\ref{fig:relative_velo_5Msun} where we plot the relative velocity between the star and NS as a function of the distance between the two stars. Each curve starts from a distance $a=8~\rsun$ and shoots inwards. Note that the starting velocities are slightly different depending on the kick angle because of the orbital velocity being added on. In a pure two-body problem without energy dissipation, the NS should reach periastron and then turn around tracing the exact same curve. For most models shown here, the NS enters the stellar interior, where orbital energy is dissipated via gravitational drag. This causes the outgoing curve to have a lower velocity than the incoming curve. In model M5-2, the deceleration is so large that the NS can never get out of the star. Model M5-0 barely touches the star, so the velocity reduction is minimal. However, it is interesting that there is any reduction at all, given that the NS was not supposed to enter the companion envelope (see Table~\ref{tab:parameters}). This highlights the importance of tidal energy dissipation in eccentric encounters. Models M5-1, M5-3 and M5-4 all plunge relatively deep into the star and out again, showing moderate velocity reduction on the way out. For models M5-3 and M5-4, the post-penetration velocities exceed the escape velocity, meaning that the binary is unbound and will become runaway stars. On the other hand, model M5-1 comes out on a bound orbit, meaning that there will be more encounters in the future. The orbital energy will be dissipated upon each encounter and should ultimately lead to a merger, forming a \ac{tzo}.

\begin{figure}
 \centering
 \includegraphics[width=\linewidth]{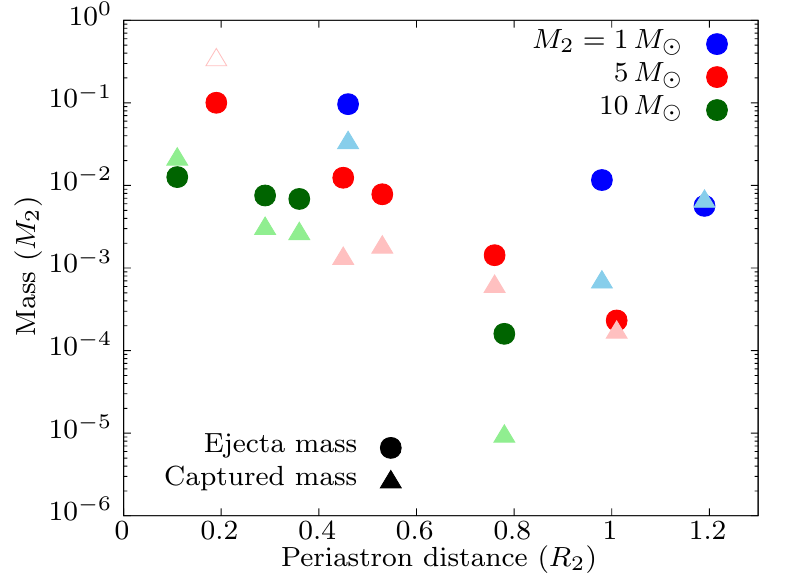}
 \caption{Dynamical ejecta and captured masses at the end of the simulation. Solid circles indicate dynamical ejecta mass whereas solid triangles indicate the mass captured by the NS. Open triangles indicate the models which immediately merge and therefore the captured mass should be treated as lower limits.\label{fig:ejecta_mass}}
\end{figure}

Figure~\ref{fig:ejecta_mass} compares final ejecta masses and captured masses as a function of periastron distance. The dynamical ejecta (circles) seem to be a fairly monotonic function despite the different models having different approach velocities. This implies that the ejecta depends more strongly on the location of the energy deposition than the amount. At most, we find $\sim10~\pc$ mass ejection in the most extreme cases. The ejecta masses are negligible ($M_\mathrm{ej}<10^{-2}M_2$) when the periastron distance is >0.6~$R_2$, meaning that envelope penetrations typically do not lead to large mass ejection.

The masses captured by the NS are less monotonic than the ejecta masses. There is an overall trend that the NS can carry away more material when it dives deeper into the star, but for example the captured mass for the second deepest model with $M_2=5~\msun$ (model M5-3) seems to be off this trend. This is likely because that model has a higher penetration velocity than the other models (see Figure~\ref{fig:relative_velo_5Msun}). The captured mass is more sensitive to the velocity at which the NS leaves the stellar surface than the depth of the penetration.

\section{Discussion}\label{sec:discussion}

\subsection{Drag force inside the star}

The amount of deceleration of the \ac{ns} through the envelope is critical in determining the final outcome. In \citet{RH21} we estimated the drag force during a penetration using the simple model by \citet{hoy39}
\begin{equation}
 f_\mathrm{drag,HL}=\frac{4\pi G^2M_\mathrm{NS}\rho}{v_\mathrm{rel}^2},
\end{equation}
where $G$ is the gravitational constant, $v_\mathrm{rel}$ is the relative velocity and $\rho$ is the local density at the location of the \ac{ns}. However, it is known that the drag force can be quite different when there are density gradients in the upstream material of the flow around the \ac{ns} \citep[]{mac15,mac17,de20}. Density gradients in main-sequence stars can be very steep, and therefore we do not expect that the drag on the \ac{ns} during envelope penetration matches the classical Hoyle-Lyttleton picture.

\begin{figure}
 \centering
 \includegraphics[width=\linewidth]{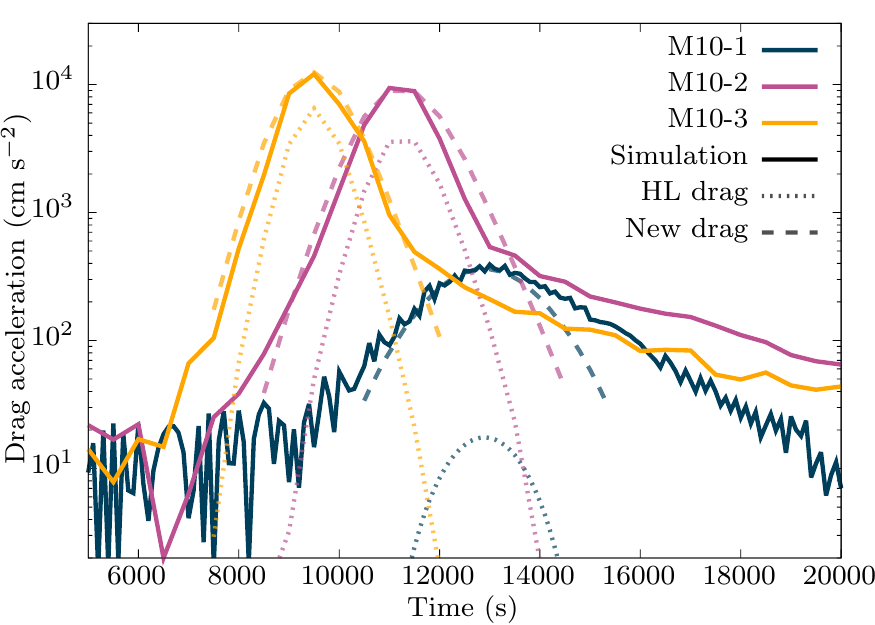}
 \caption{Time evolution of the drag force acting on the \ac{ns} in our simulations with $M_2=10~\msun$. Solid curves show the drag directly from the simulation. Dotted curves show the drag estimated from the Hoyle-Lyttleton formula using the local density. Dashed curves show the drag estimated from our new drag model for inhomogeneous flows.\label{fig:drag_timeevo}}
\end{figure}

In our simulations, we define the drag acting on the \ac{ns} by subtracting the radial gravitational acceleration from the total acceleration of the \ac{ns}. The magnitude of the radial gravitational acceleration is computed by $Gm(a)/a^2$, where $m(r)$ is the mass enclosed within a sphere of radius $r$ centred on the companion and $a$ is the distance between the \ac{ns} and centre of mass of the gas. Here we assume that the companion is undisturbed and use the initial $m(r)$ profile at all times. In Figure~\ref{fig:drag_timeevo} we plot the drag as a function of time. For the models shown, we see that the Hoyle-Lyttleton model (dotted curves) underestimates the actual drag force by a factor of a few to orders of magnitude. This is likely due to the inhomogeneity in the flow around the \ac{ns} which makes the Hoyle-Lyttleton model inapplicable. 

\begin{figure}
 \centering
 \includegraphics[width=\linewidth]{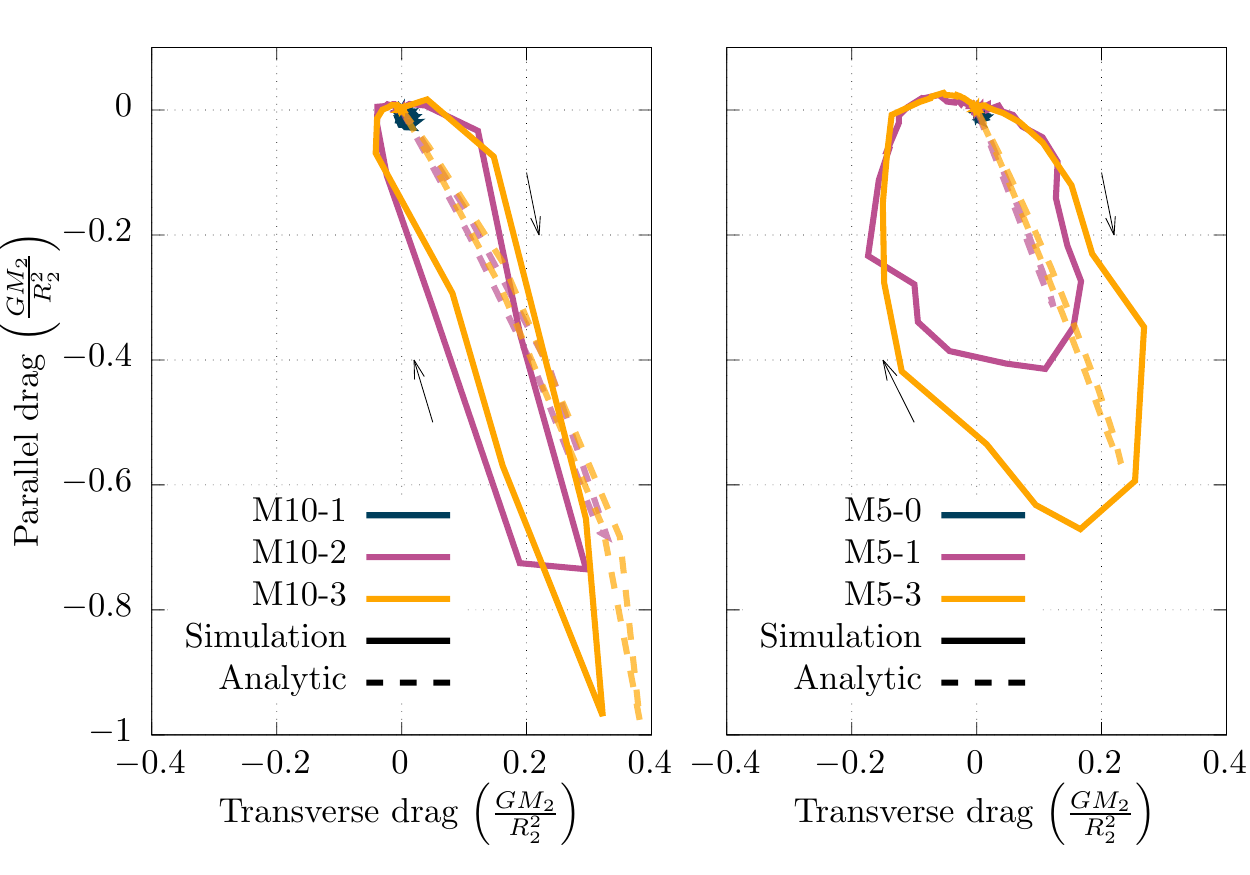}
 \caption{Direction of the drag acceleration with respect to the motion of the \ac{ns}. Positive transverse drag points towards the centre of the star whereas negative transverse drag points away. Negative parallel drag represents the acceleration anti-parallel to the motion. The drag accelerations are normalised by the acceleration at the surface of the companion star. Solid curves are the drag vectors in the simulations, while dashed curves show the drag vectors from the semi-analytic model. Black arrows indicate how the drag vector evolves in time. For clarity, we only plot the models that did not lead to immediate merger.\label{fig:drag_direction}}
\end{figure}

Another deviation from the Hoyle-Lyttleton model is the direction of the drag. When the object travels through uniform medium, the drag only acts in the direction anti-parallel to the motion due to axisymmetry. However, upstream density gradients break this symmetry, meaning that there could be transverse components to the drag. We illustrate this in Figure~\ref{fig:drag_direction} where we plot the direction of the drag with respect to the motion of the \ac{ns} (solid curves). In the classical Hoyle-Lyttleton model, the drag should only point downwards in this figure. However, we see that the drag vector has a transverse component pointing towards the higher density material which amounts to $\sim40~\pc$ of the parallel component.

\begin{figure}
 \centering
 \includegraphics[width=\linewidth]{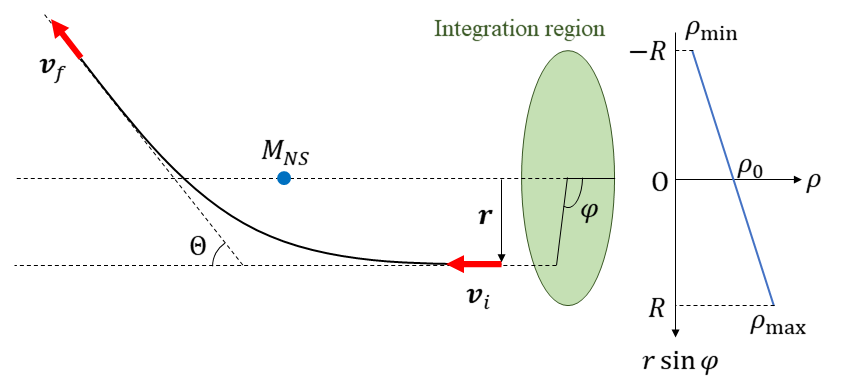}
 \caption{Notations in our analytic model for drag force in inhomogeneous medium.\label{fig:dragmodel}}
\end{figure}

In order to explain the discrepancy between the Hoyle-Lyttleton model and our simulations, we construct a new analytical model for the drag. In our model, we assume that the object is travelling supersonically with a velocity $\bm{v}_\mathrm{i}$, through gas which has a linear density gradient perpendicular to the motion (Figure~\ref{fig:dragmodel}). Each mass element of the gas that approaches the \ac{ns} is deflected by the gravity of the \ac{ns} and approaches a new velocity $\bm{v}_\mathrm{f}$. From energy conservation, the speed is unchanged $|\bm{v}_\mathrm{i}|=|\bm{v}_\mathrm{f}|$, but the direction is deflected. The difference between the incoming and outgoing velocities of the mass element is compensated by a momentum change of the \ac{ns}. We obtain the full drag by integrating this momentum change over all relevant streamlines. 

Assuming that the mass element is on a hyperbolic trajectory, the deflection angle can be computed as
\begin{equation}
 \Theta\equiv\cos^{-1}\left(\frac{\bm{v}_\mathrm{i}\cdot\bm{v}_\mathrm{f}}{|\bm{v}_\mathrm{i}||\bm{v}_\mathrm{f}|}\right)=2\tan^{-1}\left(\frac{GM_\mathrm{NS}}{rv_\mathrm{i}^2}\right),\label{eq:deflection_angle}
\end{equation}
where $r$ is the impact parameter (see Figure~\ref{fig:dragmodel}). Let us first focus on the parallel component of the drag. The parallel component of the difference between $\bm{v}_\mathrm{f}$ and $\bm{v}_\mathrm{i}$ is 
\begin{equation}
 (\bm{v}_\mathrm{f}-\bm{v}_\mathrm{i})_\|=v_\infty(1-\cos\Theta)=2v_\infty\sin^2\frac{\Theta}{2},
\end{equation}
where $v_\infty\equiv|\bm{v}_\mathrm{i}|=|\bm{v}_\mathrm{f}|$. We consider all streamlines that flow through a disk with a radius $R$ centred on the \ac{ns}. We will discuss the appropriate choice of the integration range $R$ later on. The total parallel component of the drag force on the \ac{ns} due to the deflection of all streamlines can be calculated by
\begin{equation}
 f_{\mathrm{d},\|}=\int_0^Rrdr\int_0^{2\pi}d\varphi \,\rho v_\infty\cdot2v_\infty\sin^2\frac{\Theta}{2},\label{eq:parallel_drag1}
\end{equation}
where $\rho$ is the density of the upstream material. We assume that the density has a linear gradient
\begin{equation}
 \rho(r)=\rho_0\left(1+\frac{\rho_\mathrm{max}-\rho_\mathrm{min}}{2R}r\sin\varphi\right),\label{eq:density_dist}
\end{equation}
where $\rho_0\equiv(\rho_\mathrm{max}+\rho_\mathrm{min})/2$ is the average density within the integration range. By plugging in Eq.~(\ref{eq:deflection_angle}) and Eq.~(\ref{eq:density_dist}) into Eq.~(\ref{eq:parallel_drag1}) and doing the integration, we obtain
\begin{equation}
 f_{\mathrm{d},\|}=\pi\rho_0v_\infty^2R_\mathrm{b}^2\cdot\frac12\ln\left[{\left(\frac{2R}{R_\mathrm{b}}\right)^2+1}\right],
\end{equation}
where we have defined $R_\mathrm{b}\equiv2GM_\mathrm{NS}/v_\infty^2$ as the gravitational influence radius. This expression is independent of whether the upstream density has a gradient or not, and is very similar to the Hoyle-Lyttleton model \citep[]{hoy39,ost99} when the integration range is set to $R=R_\mathrm{b}$. 

The difference between $\bm{v}_\mathrm{f}$ and $\bm{v}_\mathrm{i}$ projected on the direction of the density gradient is
\begin{equation}
 (\bm{v}_\mathrm{f}-\bm{v}_\mathrm{i})_\perp=v_\infty\sin\Theta\sin\varphi.
\end{equation}
Again we can calculate the total transverse force as
\begin{equation}
 f_{\mathrm{d},\perp}=\int_0^Rrdr\int_0^{2\pi}d\varphi \,\rho v_\infty\cdot v_\infty\sin\Theta\sin\varphi,\label{eq:transverse_drag1}
\end{equation}
and by integrating this along with Eqs.~(\ref{eq:deflection_angle}) and (\ref{eq:density_dist}) gives
\begin{equation}
 f_{\mathrm{d},\perp}=\pi(\rho_\mathrm{max}-\rho_\mathrm{min})v_\infty^2R_\mathrm{b}^2\cdot \frac{1}{16}\left[\frac{2R}{R_\mathrm{b}}-\frac{R_\mathrm{b}}{2R}\ln\left(1+\left(\frac{2R}{R_\mathrm{b}}\right)^2\right)\right].
\end{equation}
Note that this vanishes in the absence of a density gradient ($\rho_\mathrm{max}=\rho_\mathrm{min}$).

Appropriately choosing the integration radius is very important given that the transverse drag has a linear dependence on $R$ at large $R$. Some of our assumptions break down as we extend the integration range. For example, we have assumed that the upstream density distribution is planar. In reality the star is spherical, and therefore the upstream density distribution is curved, with less volume in the higher-density regions. By taking an inadequately large $R$, we will overestimate the contribution of drag from the higher-density regions. We have also ignored self-gravity of the gas, whereas the star is actually self-gravitating. In sufficiently deep parts of the star, the gravitational pull from the stellar interior dominates over the gravity from the \ac{ns}. We choose the integration radius as $R=\min\left(R_g,R_c\right)$, where $R_g$ is the radius at which the inwards gravity from the stellar interior equals the outwards force from the \ac{ns} and $R_c$ is the radius up to which the density distribution can be regarded as planar. $R_c$ must be set smaller than at least half the separation between the \ac{ns} and stellar centre. In our model, we set it as $R_c=a/4$.

To apply our model to dynamical integrations, we need a scheme to choose the relevant density and density gradients that the \ac{ns} encounters. For simplicity, we have assumed in our model that the density gradient is linear over our integration region. However, this may be a bad approximation especially when the integration range is large, since the density gradient in stars are usually more well approximated by power laws. If we simply take the density and density gradient within the star at the location of the \ac{ns}, we will greatly underestimate the density at the edge of the integration region. This is problematic, as the highest density region within the integration range has the strongest influence on the drag. To capture the relevant scales, we choose to pick out the maximum and minimum densities within the integration range and assume a linear gradient between those two values.

With our new model, we can estimate what the drag vector would be, given the relative position and velocity between the star and \ac{ns}. We plot the evolution of the drag vectors as dashed curves in Figure~\ref{fig:drag_direction}. Although not perfect, the overall direction and magnitudes are in rough agreement with the simulations. It is especially good for the higher companion mass models (left panel) where the companion star is less distorted by the interaction. The time evolution of the drag strengths are also in good agreement as shown in Figure~\ref{fig:drag_timeevo} (dashed curves). They only start to deviate at later times when the star is significantly distorted due to the interaction, which we do not take into account in our model. We caution that our semi-analytical model is not applicable for cases where the companion mass is small ($M_\mathrm{NS}\approx M_2$). This is due to the significant tidal deformation and energy dissipation before collision.

\subsection{Branching ratios of different outcomes}

With the model for the drag force provided in the previous section, we can now predict the outcome of post-SN collisions with simple dynamical integrations. We use the few-body dynamical integrator described in \citet{RH21}, and implement the drag force model as described in the previous section. We assume the pre-SN binary was on a circular orbit, and the coordinate origin is set at the centre of mass. We then instantaneously change the mass of the primary star from its initial value $M_1$ to $M_\mathrm{NS}$, assuming that the difference is lost in the ejecta and neutrino emission. At the same time, the primary star (now NS) is given a kick in its own frame and the companion star is given an impact velocity induced by ejecta-companion interaction. We estimate the impact velocity as
\begin{equation}
 v_\textrm{im}=C_\mathrm{eci}\frac{\sqrt{2E_\mathrm{exp}(M_1-M_\mathrm{NS})}}{M_2}\cdot \frac{1-\sqrt{1-(R_2/a)^2}}{2},
\end{equation}
where $E_\mathrm{exp}$ is the explosion energy, which we set to a canonical value $10^{51}$~erg. The momentum transfer efficiency $C_\mathrm{eci}$ is set to $1/3$, based on results from hydrodynamical simulations of ejecta-companion interaction \citep[]{RH18}. The impact velocity is applied impulsively, directed away from the primary star. With these initial conditions, we integrate the orbit of the two stars with a drag force whenever the two stars get closer than $R_2$.

\begin{figure}
 \centering
 \includegraphics[width=\linewidth]{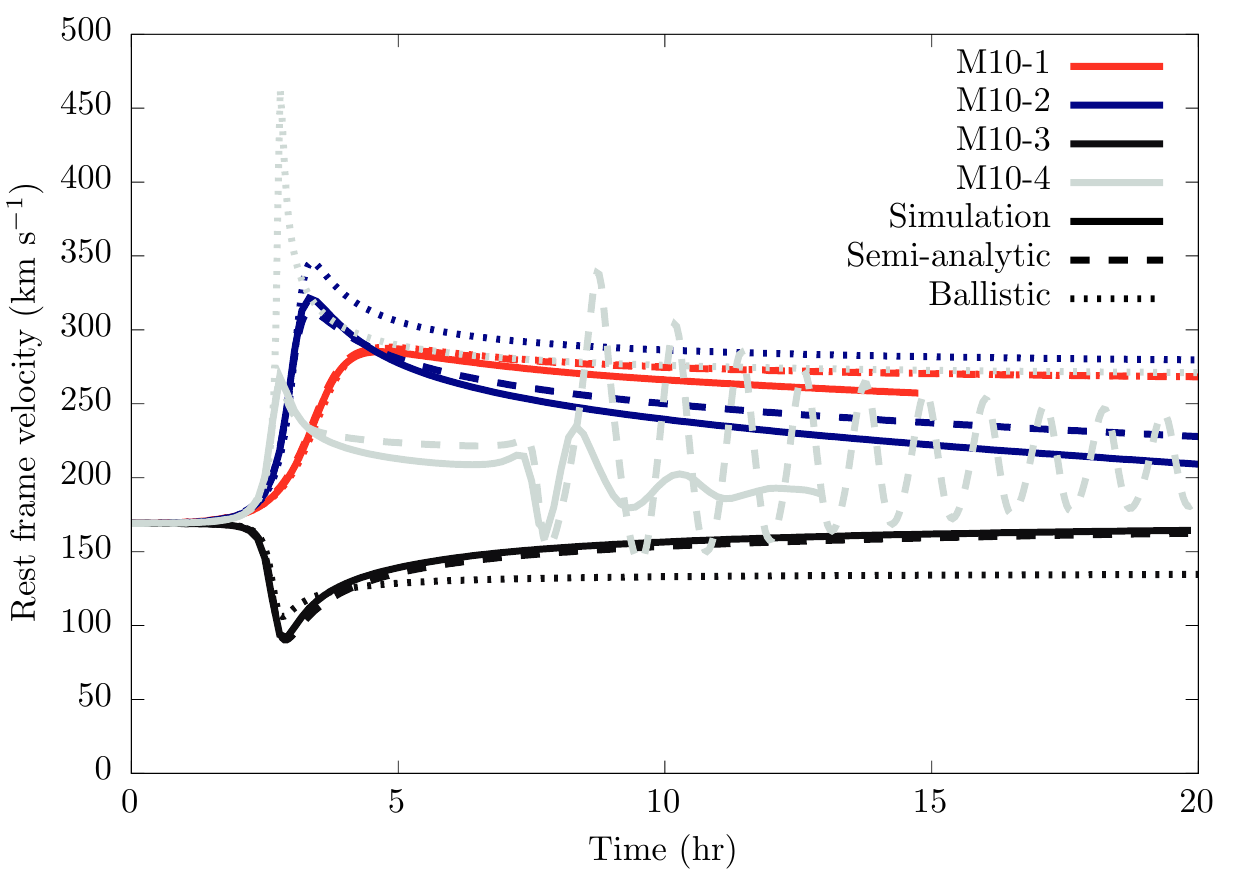}
 \caption{Rest frame velocities of the companion star as a function of time. Solid curves show results from the hydrodynamical simulations whereas the dashed curves show results from our semi-analytical integration. Dotted curves show the results of a ballistic trajectory without drag forces.\label{fig:semiana_velo}}
\end{figure}

In Figure~\ref{fig:semiana_velo} we show some examples of the semi-analytical integration. Here we set $v_\mathrm{im}=0$ to be consistent with the hydrodynamic simulations. For models M10-1, M10-2 and M10-3, we see that the hydrodynamic simulations (solid curves) and semi-analytical models (dashed curves) agree fairly well. These models reach moderate depths within the star at periastron (see Table~\ref{tab:parameters}) and therefore have a moderate deviation from ballistic trajectories (dotted curves). Even for model M10-4, which leads to an immediate merger, the semi-analytic model predicts the deceleration upon the first encounter remarkably well (first peak in the grey curve). After the second periastron, the star is already significantly distorted so many of our assumptions break down and thus the semi-analytic model starts to largely deviate from the hydrodynamic simulations.

These results demonstrate that the semi-analytic models can predict the trajectories at least upon the first encounter quite accurately. We now perform the same dynamical integration for all kick angles. We classify the results into 5 different cases: (i) binary is disrupted without any collision, (ii) binary survives but the two stars do not collide, (iii) the NS penetrates through the companion envelope on an unbound orbit, (iv) the NS penetrates through the companion envelope on a bound orbit, (v) the NS never emerges out of the companion after collision and immediately merges. 

\begin{figure}
 \centering
 \includegraphics[width=\linewidth]{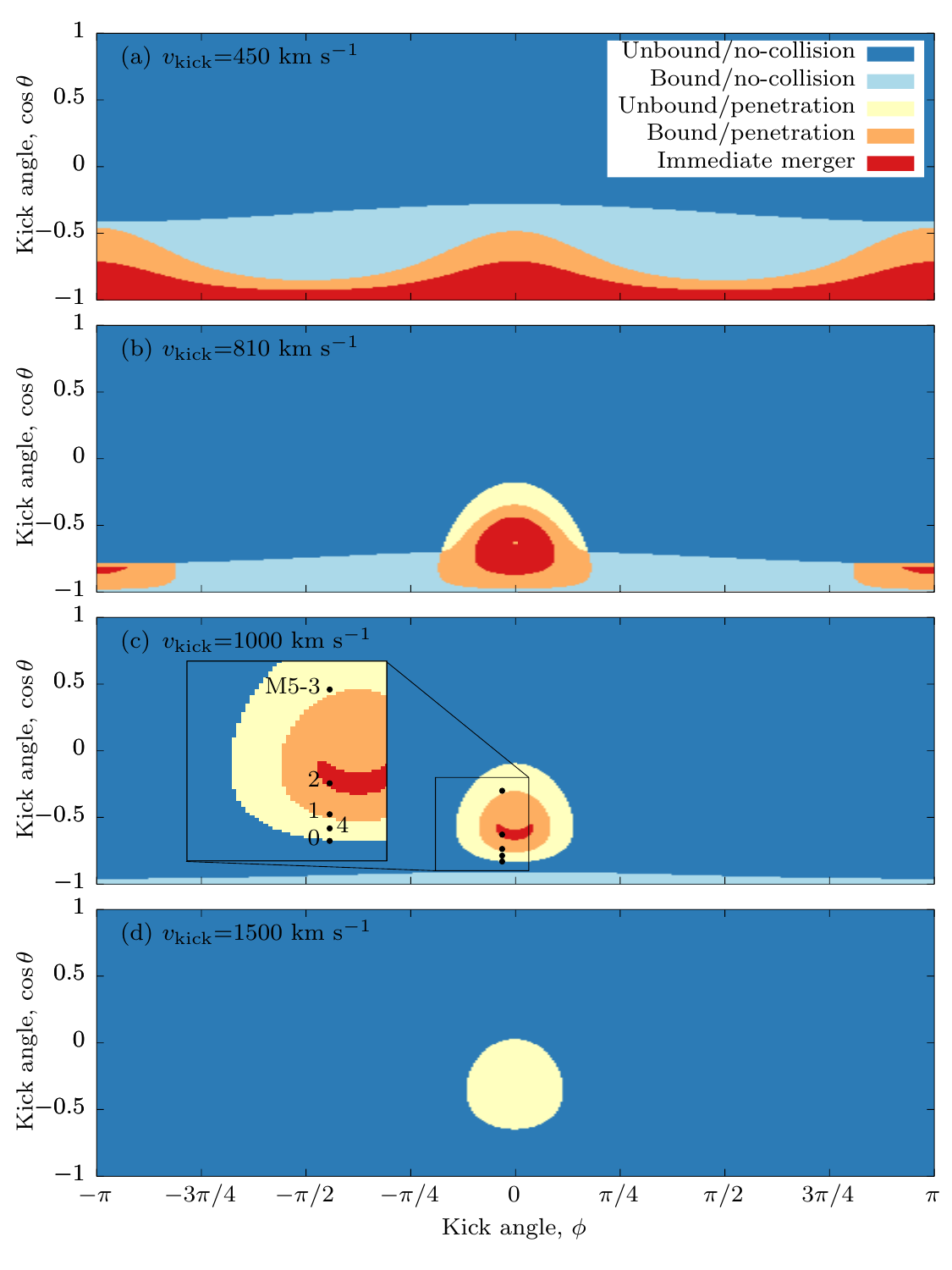}
 \caption{Diagrams showing the expected outcome depending on the kick direction. Each colour represents a different outcome as described in the legend. Model parameters are $M_1=6~\msun$, $M_2=5~\msun$, $M_\mathrm{NS}=1.4~\msun$, $a=8~\rsun$ and the kick magnitudes are labelled in each panel. The black dots in panel (c) mark the models chosen for our hydrodynamical simulations. The numbers correspond to the model number as defined in Table~\ref{tab:parameters}.\label{fig:ocdiagram_5Msun}}
\end{figure}

In Figure~\ref{fig:ocdiagram_5Msun} we show a diagram that depicts the outcomes as a function of kick angle for a given system with $M_1=6~\msun$, $M_2=5~\msun$, $M_\mathrm{NS}=1.4~\msun$ and $a=8~\rsun$. In the absence of a kick ($v_\mathrm{kick}=0$), this system will stay bound without any collision (the whole plane will be light blue). When the kick magnitude is comparable to the orbital velocity, it becomes possible to direct the newborn NS towards the companion and collide with it (panel (a)). When the kick is strong enough to collide with the NS on hyperbolic trajectories, we start to see cases where the NS penetrates the companion envelope on unbound orbits (yellow region in panels (b) and (c)). Depending on how deep the NS plunges in, the NS can also penetrate the envelope on bound orbits (orange regions) or never emerge from the surface again (red regions). In the former case, the NS will plunge into the companion multiple times before eventually dissipating enough energy to merge with the star. At sufficiently high kick magnitudes, the NS can plunge through the star no matter how deep it is directed (panel (d)).

\begin{figure}
 \centering
 \includegraphics[width=\linewidth]{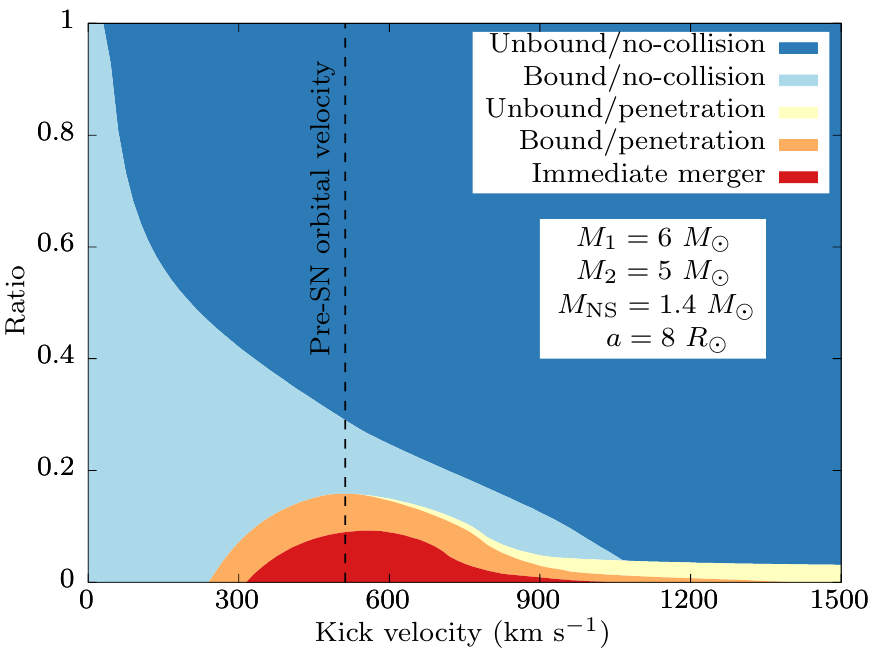}
 \caption{Branching ratios of the different outcomes as a function of the kick velocity. The vertical dashed line indicates the pre-SN orbital velocity as a visual guide.\label{fig:branching_5Msun}}
\end{figure}

In Figure~\ref{fig:branching_5Msun}, we show the branching ratios of each outcome as a function of the kick magnitude. Here we assume that the kick is isotropically distributed over all angles. The collisions occur in the yellow/orange/red regions. We can see that the collision probability can reach as high as $\sim16~\pc$ at optimal kick velocities ($v_\mathrm{kick}=v_\mathrm{orb}$, where $v_\mathrm{orb}$ is the pre-SN orbital velocity). This maximum fraction strongly depends on the pre-\ac{sn} orbital separation and can reach up to $\sim25~\pc$ in optimal situations. Among the colliding cases, roughly less than half lead to immediate mergers and the rest cause envelope penetrations. Unbound penetrations start to appear when the kick velocity exceeds the pre-SN orbital velocity. 

\section{Applications}\label{sec:application}

\subsection{Hypervelocity stars}
Hypervelocity stars are a particular class of runaway stars that have velocities exceeding that of the local escape velocity of the Galaxy (see \citealt{bro15} for a review on hypervelocity stars). There are two main formation mechanisms that are usually discussed in the literature that both involve disruption of a binary, either by the tidal field of a supermassive black hole \citep[``Hills mechanism'';][]{hil88} or through a SN explosion \citep[``Blaauw mechanism'';][]{bla61}\footnote{In some studies, the stars ejected through the Hills mechanisms are called hypervelocity stars while the stars ejected through the Blaauw mechanism are called runaway stars. In this paper, we use the term more generally and call all escaping stars as hypervelocity stars.}. Other channels involve ejection via dynamical interactions in clusters \citep[]{pov67,leo91}. The Hills mechanism generally predicts higher velocities than the Blaauw mechanism or dynamical ejection channels, and therefore are considered to dominate the number of hypervelocity stars \citep[]{bro09,per12}. However, there is growing evidence of a number of hypervelocity stars originating from the Galactic disk instead of the centre, which cannot be produced by the Hills mechanism \citep[]{heb08,irr10,irr18a,irr19}.

The ejection velocity of stars after disruption of binaries is determined by a combination of the pre-SN orbital velocity and the NS natal kick magnitude and direction. For the majority of cases, the companion star will travel at its pre-SN orbital velocity, which typically does not exceed $\sim200$--$300$~\kms for B-type stars. Only in very rare cases where the NS kick velocity is high and directed close to the companion star surface, the companion experiences a swing-by effect and can be accelerated to higher velocities. However, even in the most favourable conditions, the ejection velocity for B-type stars cannot exceed a threshold of about $540$~\kms \citep[]{tau15}. The observed ejection velocities of some disk-origin B-type hypervelocity stars reach up to $\gtrsim600$~\kms, seemingly exceeding this hard upper limit \citep[]{irr18a,irr19}. We shall call these stars as the ``impossible'' hypervelocity stars in this section.

\begin{figure}
 \centering
 \includegraphics[width=\linewidth]{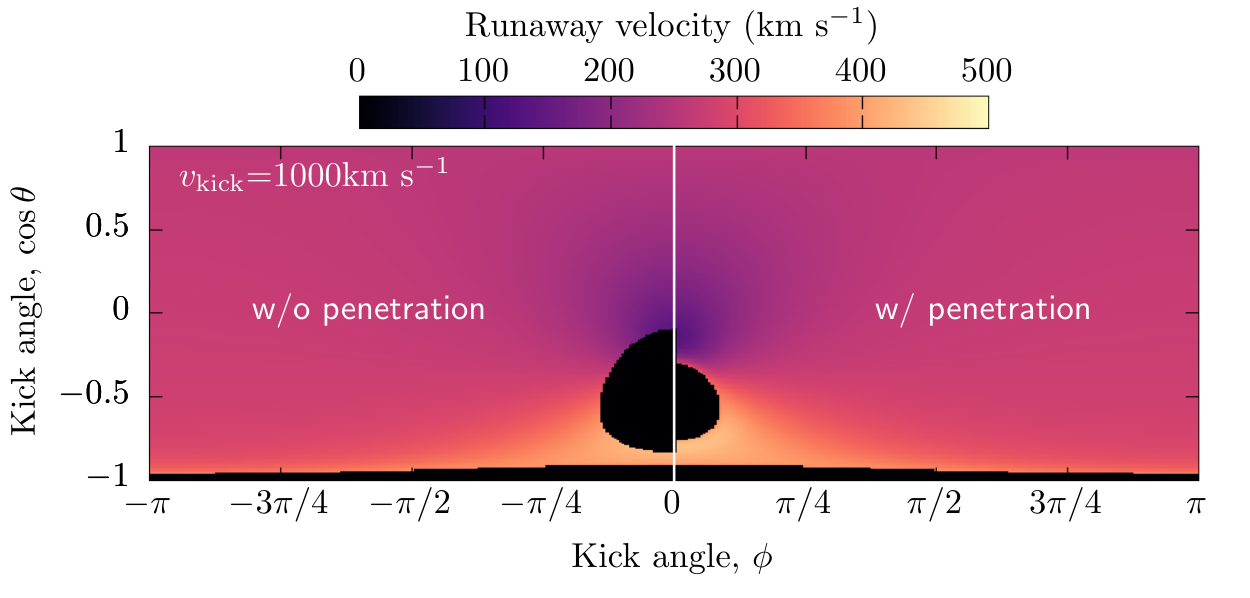}
 \caption{Runaway velocity of the companion as a function of kick direction. Model parameters are $M_1=6~\msun$, $M_2=5~\msun$, $M_\mathrm{NS}=1.4~\msun$, $a=8~\rsun$ and $v_\mathrm{kick}=1000$~\kms. The left side is computed with the same procedure as in \citet{tau98}. The right side is results from our semi-analytical calculations. Black areas show kick directions where the binary does not disrupt.\label{fig:runaway_velocity}}
\end{figure}

The main reason why there is a hard upper limit to the ejection velocity in the Blaauw mechanism is the assumption that the stars will merge if the NS is kicked into the companion. Due to this assumption, the maximum achievable velocity is restricted roughly by the surface escape velocity of the companion. We can see this in Figure~\ref{fig:runaway_velocity}. The left side of the panel shows the asymptotic runaway velocity of the companion using the analytical formulae derived in \citet{tau98}. The thin black region at the bottom shows where the binary survives the SN without getting disrupted. The circular black region in the middle shows where the NS collides with the companion and is therefore assumed to merge. The highest runaway velocities are achieved just outside the blacked-out region when the NS skims the surface of the companion. The right side show results taken directly from our semi-analytical simulations. Here, envelope penetration can occur and thus the central blacked-out region is smaller compared to the left side. Consequently, the system has a slightly higher chance of creating runaway stars. Also, the maximum achievable runaway velocity is slightly higher because the stars can be scattered on a closer orbit, applying a stronger swing-by effect.

\begin{figure}
 \centering
 \includegraphics[width=\linewidth]{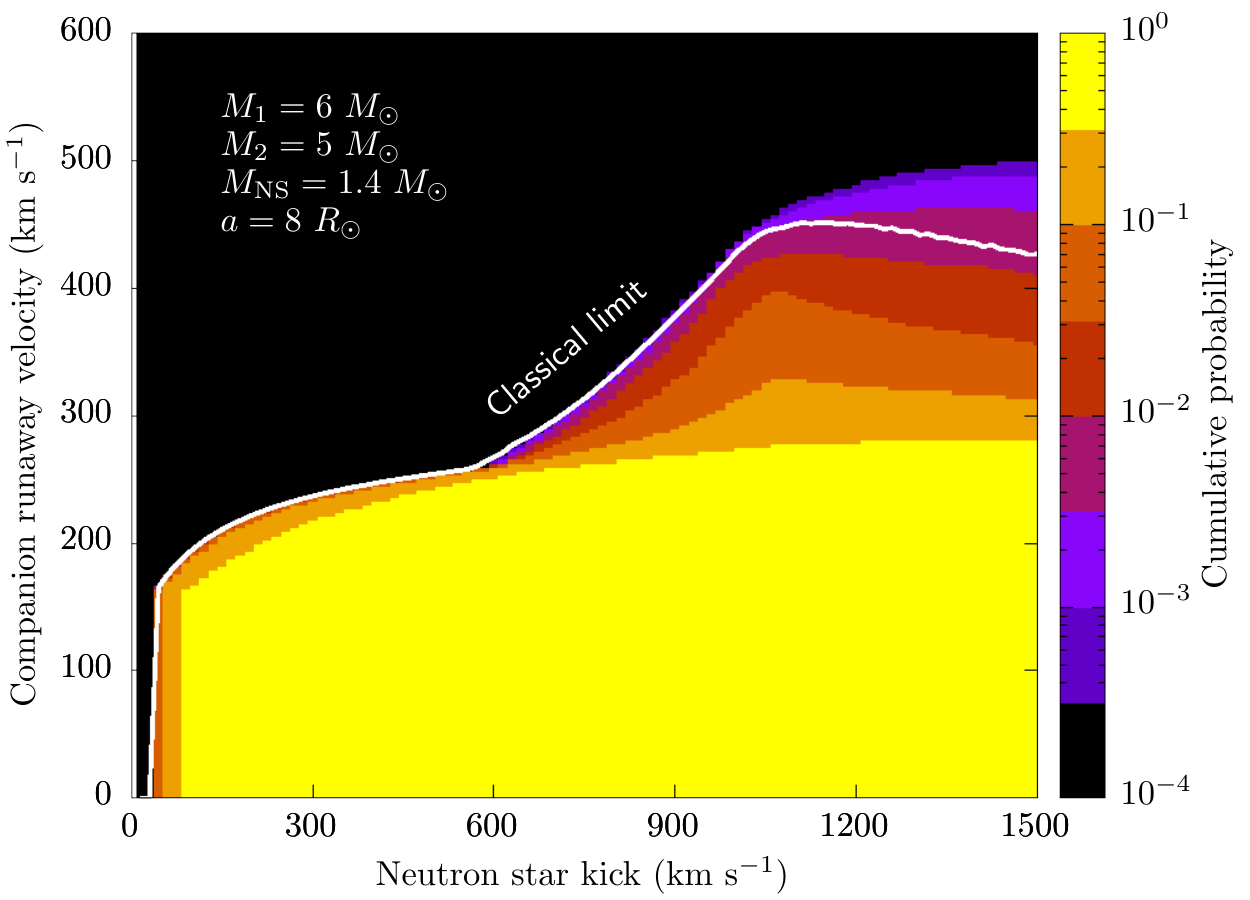}
 \caption{Probability distribution of runaway velocities for various kick magnitudes. Each vertical slice shows the cumulative probability distribution of the runaway velocity from top to bottom. In other words, the colours at each position indicate the probability of having a runaway velocity above that value given a NS kick velocity. The white curve shows the classical upper limit using the same approach as in \citet{tau15}.\label{fig:runaway_grid}}
\end{figure}

Based on these calculations, we can derive the probability distributions of runaway velocities given a \ac{ns} kick velocity. We show cumulative probability distributions over a range of \ac{ns} kick velocities in Figure~\ref{fig:runaway_grid}. The white curve shows the maximum achievable runaway velocity in classical studies where envelope penetrations are not taken into account \citep[]{tau15}. There is an optimal \ac{ns} kick velocity around $v_\mathrm{kick}\sim1000$~\kms where the upper limit reaches a maximum. In our calculations, the maximum achievable velocity perfectly agrees with the classical limit up to a certain point. However, at extremely high NS kick velocities ($v_\mathrm{kick}\gtrsim1000$~\kms), the maximum achievable velocity exceeds the classical limit due to unbound envelope penetrations. As a result, we achieve a new theoretical upper limit that is higher than the classical limit by $\sim10~\pc$. \citet{tau15} derived that in the most optimal situations ($M_1=5~\msun, M_2=3.5~\msun, a=6~\rsun, E_\mathrm{exp}=1.23\times10^{51}~\textrm{erg}$), the runaway velocity can reach up to $\sim540$~\kms. With our envelope penetration, we can now reach up to $\sim600$~\kms with the Blaauw mechanism, reaching the territory of the impossible hypervelocity stars.

The probability of exceeding the classical limit is only $\sim0.5~\pc$, given that the binary orbit is very tight and the NS kick is fast enough. To derive the rate of impossible hypervelocity star formation, this needs to be multiplied by the probability of having a binary satisfying the necessary conditions. Binary population synthesis studies predict that $\sim70~\pc$ of stripped-envelope SNe should have a main-sequence companion at the point of SN \citep[]{zap17}. Most of these systems are expected to have relatively large orbital separations unless they had experienced common-envelope evolution, which remains one of the most uncertain aspects of binary evolution (see \citealt{iva20} and references therein). Although very uncertain, \citet{ren19} find that $\sim1~\pc$ of systems could have pre-SN orbital separations close enough to create runaway stars. We take $\sim1~\pc$ as a rough estimate of the fraction of systems that have tight enough orbits where envelope penetration is possible. Alternatively, we can provide a very optimistic estimate assuming that most stripped-envelope \ac{sn} progenitors are formed through some form of binary interaction, therefore having tight enough orbits. The number of stripped-envelope \acp{sn} is roughly $\sim1/3$ of the entire population of \acp{ccsn} \citep[]{li11,smi11}. As for the kicks, the fraction of NSs born with velocities $v_\mathrm{kick}>1000$~\kms is a few$~\pc$ assuming a kick velocity distribution similar to \citet{ver17}. Combining these numbers, we roughly obtain a formation rate for impossible hypervelocity stars as $\sim10^{-6}$--$10^{-5}\times$~\ac{ccsn} rate. Taking a typical \ac{ccsn} rate ($\sim0.02$~yr$^{-1}$) and main-sequence lifetime of B-stars ($\sim10^8$~yr), there should be roughly $\sim1$--$10$ impossible hypervelocity stars in our Galaxy.

In summary, our new binary disruption model with envelope penetrations can in principle exceed the classical theoretical upper limit for runaway stars, but only by $\sim10~\pc$. In the most optimal situations, the maximum achievable runaway velocity for B-stars can reach $\sim600$~\kms, which is enough to explain the origin of some impossible B-type hypervelocity stars \citep[]{irr18a,irr19}. However, further investigation is required involving rigorous binary population synthesis to provide a more accurate formation rate.

\subsection{Pulsar planet formation}
Pulsar planets are planets orbiting pulsars, which was actually the first exoplanets to be discovered \citep[]{wol92}. So far, there have been only about half a dozen firm detections of pulsars hosting planets \citep[]{nit22}. The origin of these planets remain elusive, with several scenarios being proposed, such as dynamical capture from other stars, forming proto-planetary disks through fall-back, stellar mergers or tidal disruption, and evaporating stellar companions into planets \citep[]{pod93,phi93}. The properties of pulsar planets have a wide diversity, with different masses, orbital periods and eccentricities so it is likely that there are multiple contributing formation channels.

Of particular importance is PSR B1257+12, which hosts three planets with masses $0.02, 4.3, 3.9~M_\oplus$ at orbital separations $0.19, 0.36, 0.46$~AU \citep[]{kon03}. The eccentricity of each orbit is extremely small ($e<0.025$), meaning it was more likely produced through a proto-planetary disk rather than a dynamical capture scenario. Some NSs are now being discovered with debris disks around it, providing strong support for the channels involving proto-planetary disk formation \citep[]{wan06}.

In our scenario, the NS captures and carries away part of the companion's mass as it penetrates the envelope. This captured mass naturally has some angular momentum so it eventually forms a partially rotationally supported torus towards the end of our simulation. It may be possible that this material eventually cools to form a proto-planetary disk and start forming planets. Our simulations show that up to $\lesssim3\times10^{-2}~\msun\approx30~M_J$ of mass can be carried away by the NS (Figure~\ref{fig:ejecta_mass}), which exceeds the mass of the most massive pulsar planets observed. If we assume a metallicity of $Z=0.014$, the captured matter contains up to $\lesssim0.4~M_J\approx130M_\oplus$ of heavy elements, enough to form several rocky super-Earth planets.

\begin{figure}
 \centering
 \includegraphics[width=\linewidth]{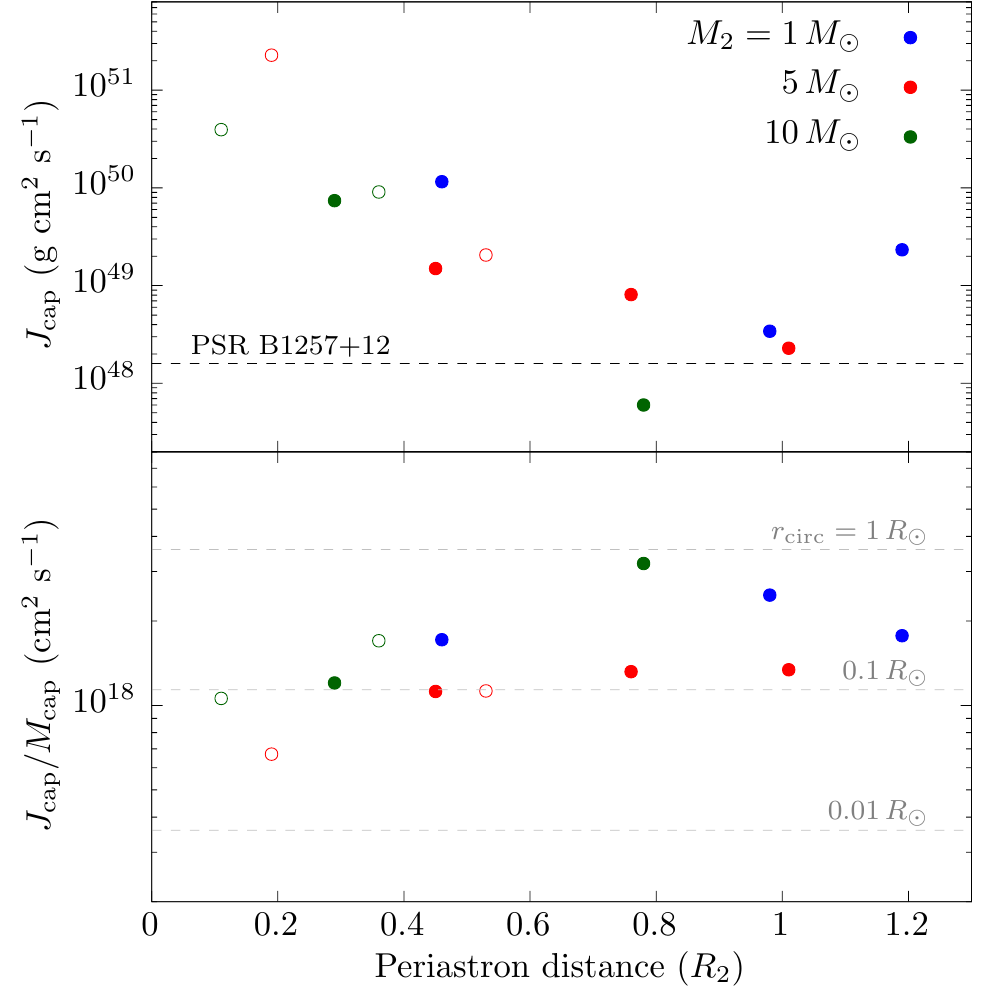}
 \caption{Angular momentum of the material carried away by the NS. Colours of the plots distinguish the companion star model used in the simulations. Open circles are models that do not lead to unbound penetrations. The top panel shows the total captured angular momentum and the bottom panel shows the specific angular momentum. The black dashed line in the top panel indicates the total angular momentum in the planetary system around PSR B1257+12. Grey dashed lines in the bottom panel indicate the specific angular momentum of a particle orbiting the NS at an orbital radius $r_\mathrm{circ}$.\label{fig:captured_AM}}
\end{figure}

The angular momentum in the captured material ($J_\mathrm{cap}$) are shown in Figure~\ref{fig:captured_AM}. We can see that the total captured angular momentum decreases with periastron distance (top panel). It is roughly proportional to the amount of captured mass, and thus the average specific angular momentum lie in a relatively narrow range (bottom panel). If all of the captured matter condense into a single planet, the circularization radius would be around $\sim0.1$--$1~\rsun$, which is much tighter than most observed pulsar planets. However, for most of our models, the total amount of captured angular momentum is sufficiently higher than the total angular momentum in PSR B1257+12. Depending on how the angular momentum is re-distributed as the captured material cools and forms into a disk, it may be possible to form planetary systems similar to PSR B1257+12.

One of the benefits of our scenario is that the accreted material naturally has a large angular momentum reservoir compared to some other accretion scenarios such as \ac{sn} ejecta fallback \citep[]{bai91}. With lower mass companions, our scenario is equivalent to the tidal disruption scenario, but our scenario generally works better at the high-mass limits where there can be a larger amount of mass captured.

The favourable conditions for our scenario to work is to have more massive companions, larger NS kicks and relatively low impact parameters (=small periastron distances). In particular, the kick velocity has to be at least greater than the pre-SN orbital velocity in order to have a chance of experiencing unbound penetrations (yellow regions in Figures~\ref{fig:ocdiagram_5Msun} and \ref{fig:branching_5Msun}). The required conditions overlap with the conditions to create hypervelocity stars. In such cases, the NS actually ends up with a rather small rest frame velocity despite being given a large kick because most of the momentum is passed onto the companion through the swing-by effect. This implies that most of the planet-forming pulsars will be retained in the Galaxy without being ejected. The favourable parameter space may be further expanded if we take into account the inflation/ablation of the envelope through the \ac{sn} impact \citep[]{nak91,oga21}.

The expected occurrence rate of this scenario is larger than the hypervelocity star rate as we do not require the kick velocity to be extremely fast. Following the same arguments as the previous section, we take $\sim1$--$30~\pc$ as a rough estimate of the fraction of systems that have orbits tight enough for \ac{ns}--companion collisions to occur. Next, the NS needs to be given a kick velocity above that of the pre-SN orbital velocity. Again assuming a simple kick velocity distribution similar to that of \citet{ver17}, roughly $\sim50~\pc$ of NSs will be given kicks greater than $400$~\kms. From Figure~\ref{fig:branching_5Msun}, the probability of having unbound penetrations is a few per cent as long as the kick velocity is greater than the orbital velocity. Combining all the above, we roughly predict that $\sim10^{-5}$--$10^{-3}$ of NSs should have planetary systems similar to PSR B1257+12. Recent surveys place an upper limit on planet-hosting NSs to be no more than $0.5~\pc$ \citep[]{nit22}. Our estimated rate does not conflict with the observed upper limit, but is considerably lower. Future surveys with a larger sample of pulsars may be able to test our scenario.

\subsection{Peculiar supernovae}
The primary target for collisions to occur are very tight binaries that cause stripped-envelope SNe. In particular, we expect it to occur in hydrogen-deficient SNe classified as type Ib or Ic. The collision and penetration both occur on a few hour timescale, so given that typical type Ib/c SNe are bright for weeks to months, the whole interaction process will be well embedded inside the SN photosphere. This makes it improbable to observe the collision itself.

Our simulations show that upon collision, up to $\lesssim0.1~M_2$ of matter can be ejected from the system and if the NS can penetrate the companion, it carries away up to $\lesssim0.01~M_2$ (Figure~\ref{fig:ejecta_mass}). As the captured matter settles around the NS, the inner parts could cause direct accretion onto the NS. It is not possible to obtain a mass accretion rate from our simulations as we do not resolve the dynamics in the vinicinity of the NS, but it is likely to be highly super-Eddington. If so, the accretion will be radiatively inefficient and could cause outflows or even jets \citep[e.g.][]{nar94}. The outflow can then interact with the inner parts of the \ac{sn} ejecta, serving as an extra energy source to power the light curve, analogous to fallback \citep[e.g.][]{dex13}, magnetar spin-down \citep[e.g.][]{ost71}, and jet-powered scenarios \citep[e.g.][]{sok16}. This could be a viable channel for producing hydrogen-deficient superluminous \acp{sn} \citep[]{qui11}. Even if only $\sim10~\pc$ of the $0.01~\msun$ of captured matter is accreted ($M_\mathrm{acc}\sim10^{-3}~\msun$), there would be roughly $E_\mathrm{acc}\sim M_\mathrm{acc}c^2\sim 10^{51}$~erg of accretion energy available, which is sufficient to produce superluminous \acp{sn}.

According to our modelling, the majority of envelope penetrations would be bound penetrations (Figure~\ref{fig:branching_5Msun}), meaning that the NS would eventually come back to interact with the companion again and possibly penetrate multiple times. Each time, the \ac{ns} could carry away extra envelope matter, refueling itself to power more outflows. This could lead to multiple powering events to the SN, causing bumps in the light curve. The interval between the first and second penetrations strongly depends on the kick magnitude and direction, and can take any value up to a few months. The subsequent intervals will rapidly decrease, as the interaction velocity goes down and is more efficiently decelerated each time. If the accretion onto the \ac{ns} can be sufficiently prolonged, or if the second interaction occurs late enough when the \ac{sn} ejecta have become optically thin, the accretion itself may become observable at late times in X-rays or radio.

Late-time bumps have actually been discovered in recent superluminous SN light curves such as in SN2015bn \citep[]{nic16} and SN2019stc \citep[]{gom21}. In fact, it appears to be a ubiquitous property of hydrogen-deficient superluminous SNe \citep[]{hos22}. Common explanations for these bumps are interaction with inhomogeneous circumstellar medium \citep[e.g.][]{sor16}, variable spin-down of magnetars \citep[]{chu22,mor22}, or jets powered by accretion from (inflated) companion stars \citep[]{sok22,hob22}. Our new scenario is similar to the magnetar and jet-powered scenarios as it is heating the ejecta from inside, whereas circumstellar matter interaction occurs on the outer parts of the ejecta. Unlike the magnetar scenario, our scenario is tapping on accretion energy rather than rotational energy, so it does not require rapidly rotating NSs. In that sense our scenario is similar to the jet-powered scenario except that the accretion is caused by direct penetration rather than gentle capture from envelopes inflated by the SN ejecta \citep[]{RH18,oga21}. Also, the system should end up as a \ac{tzo} while the jets scenario would predict a survived binary.

\subsection{Thorne-\.Zytkow objects}
There are several competing ideas as to how to form \acp{tzo}. The classical scenario is the merger between donors and NSs in massive X-ray binaries \citep[]{taa78}, which is considered to dominate the formation rate \citep[]{pod95a}. Another major scenario is the one we focus on in this paper, which is the direct collision of NSs and companions following SNe \citep[]{leo94}. Other likely subdominant scenarios are direct collisions between NSs and stars in dense clusters \citep[]{ray87} and merger-induced collapse of oxygen-neon-magnesium white dwarfs inside common envelopes \citep[]{abl22}.

Out of the five different outcomes we have identified, the immediate mergers and bound penetrations will inevitably lead to the formation of \acp{tzo}. Additionally, the non-colliding cases could also later form \acp{tzo} if it is on a bound orbit with a sufficiently short periapsis. The tidal interactions at periastron could efficiently dissipate the orbital energy and cause a delayed merger. Roughly $\sim1/4$ of all bound systems with massive star companions are expected to form \acp{tzo} in this way \citep[]{bra95}.

Our simulations show that unbound penetrations are rather rare, and can only occur when the NS kick is high enough. All other collisions will either immediately merge or merge after a few envelope penetrations. Therefore, accounting for envelope penetrations should not have a large influence on the predicted formation rate of \acp{tzo} through this channel, which is about $\sim10^{-4}$~yr$^{-1}$ \citep[]{pod95a}. 

\acp{tzo} are expected to be virtually indistinguishable from red supergiants. The only characteristics are peculiar chemical abundances \citep[]{bie91,bie94,can92,pod95a} or they could emit detectable gravitational waves \citep[]{dem21}. Later on, they could cause bright explosions once they run out of fuel \citep[]{mor18,mor21}. While there are several candidate objects \citep[]{lev14,bea18}, there are no conclusive detections of \acp{tzo} yet so it remains to be seen whether the theoretical rate predictions are correct.

In our hydrodynamical models with $M_2=1~\msun$, we eject up to $\sim0.1~\msun$ from the system while only $\sim0.03~\msun$ is bound to the NS (Figure~\ref{fig:ejecta_mass}). That is with a relatively extreme NS kick velocity $v_\mathrm{kick}=1000$~\kms. When we compare models M1-1 with M1-3, M1-1 has an order of magnitude higher captured mass despite the impact parameter being larger. This is because in M1-1, the kick is cancelled more by the orbit, leading to a lower approach velocity. Hence it suggests that with lower more typical kick velocities, the captured mass can substantially increase, forming \acp{tzo}. Such low-mass \acp{tzo} are expected to be supported by accretion rather than nuclear burning \citep[]{can92}, so the surface abundance may be less peculiar.

\section{Summary and conclusion}\label{sec:conclusion}
We performed 3D hydrodynamical simulations of new-born \acp{ns} colliding with binary companions after \ac{sn} explosions. We focused on cases when the \ac{ns} kick velocity is large ($v_\mathrm{NS}=1000$~\kms) and explored two parameter variations: companion mass and kick direction. Our results can be summarized as follows.
\begin{enumerate}
 \item When the companion mass is smaller than the \ac{ns} ($M_2<M_\mathrm{NS}$), the collision shows dynamics similar to partial tidal disruption events, causing some mass ejection as well as accretion onto the \ac{ns}.
 \item When the companion mass is greater than the \ac{ns} ($M_2>M_\mathrm{NS}$), the collision can lead to three different outcomes: immediate merger, bound penetrations and unbound penetrations.
 \item When the \ac{ns} penetrates the companion's envelope, it can carry away up to a few $~\pc$ of the envelope mass.
\end{enumerate}

We constructed a semi-analytical formula to describe the drag force inside stars with steep density gradients. After checking that our formula agrees well withour hydrodynamical simulations, we used that to integrate the orbits of \ac{ns}--star collisions over a much wider parameter space. We find that envelope penetrations are not a rare phenomenon and can occur with even moderate kick velocities. Most cases are bound penetrations, where it is expected that there will be multiple penetration events before finally merging. Unbound penetrations can only occur when the kick velocity is greater than the pre-SN orbital velocity.

We speculate on four different kinds of objects and phenomena that could be associated with \ac{ns}--star collisions.
\begin{description}
 \item[\textit{Hypervelocity stars}] The unbound penetration cases can sometimes accelerate the companion star to very high velocities. In the most optimal situations the runaway velocity can exceed the classical theoretical upper limit by $\sim10~\pc$, up to $\sim600$~\kms. This small extra gain may be sufficient to explain the origin of some observed B-type hypervelocity stars that exceed the classical theoretical upper limit. We roughly expect about $\sim1$--$10$ such stars to exist in our Galaxy.

 \item[\textit{Pulsar planets}] The matter carried away by the \ac{ns} after unbound penetrations contain a large amount of angular momentum. Both the mass and angular momentum is sufficient to create protoplanetary disks that may eventually form planetary systems around pulsars, similar to PSR B1257+12. The expected occurrence rate is rather small ($\sim4\times10^{-5}$ of all pulsars), which is consistent with the upper limits from recent surveys \citep[]{nit22}.

 \item[\textit{Peculiar supernovae}] The carried away matter can cause radiatively inefficient accretion onto the \ac{ns}, possibly forming outflows and jets. This will interact with the inner parts of the \ac{sn} ejecta and could provide an extra energy source to power the light curve. If $\sim10~\pc$ of the captured matter can accrete onto the \ac{ns}, there will be $\lesssim10^{51}$~erg of available accretion energy, which is sufficient to power the light curve up to levels of superluminous \acp{sn}. If there are multiple envelope penetrations, the heating can occur episodically, creating bumps in the light curve similar to those in SN2015bn and SN2019stn.

 \item[\textit{Thorne-\.Zytkow objects}] The majority of the collisions are either bound penetrations or immediate mergers, which means that it will form \acp{tzo}. We find that unbound penetrations are rare enough that it will not influence previous estimates of \ac{tzo} formation rates.
\end{description}

There are many possible directions for future investigation. More simulations over a wider grid of parameters over longer time-scales should be carried out to quantify the effects we have ignored in this study, such as envelope inflation by the impact of the \ac{sn}, initial rotation of the companion and subsequent collisions in bound penetration cases. We have only speculated on the possibility of powering \ac{sn} light curves with accretion onto the \ac{ns}. Radiation hydrodynamics simulations are required to model the light curves expected from our scenario. Our estimated rates for each of our outcomes are based on rough optimistic assumptions. The rates could be refined with a more rigorous binary population synthesis study.

\section*{Acknowledgements}
RH thanks Ilya Mandel and Evgeni Grishin for useful discussions on the analytical drag formula. 
This work was performed on the OzSTAR national facility at Swinburne University of Technology. The OzSTAR program receives funding in part from the Astronomy National Collaborative Research Infrastructure Strategy (NCRIS) allocation provided by the Australian Government.

\section*{Data Availability}
The data underlying this article will be shared on reasonable request to the corresponding author.

%%%%%%%%%%%%%%%%%%%%%%%%%%%%%%%%%%%%%%%%%%%%%%%%%%



\bibliographystyle{mnras}

\begin{thebibliography}{}
\makeatletter
\relax
\def\mn@urlcharsother{\let\do\@makeother \do\$\do\&\do\#\do\^\do\_\do\%\do\~}
\def\mn@doi{\begingroup\mn@urlcharsother \@ifnextchar [ {\mn@doi@}
  {\mn@doi@[]}}
\def\mn@doi@[#1]#2{\def\@tempa{#1}\ifx\@tempa\@empty \href
  {http://dx.doi.org/#2} {doi:#2}\else \href {http://dx.doi.org/#2} {#1}\fi
  \endgroup}
\def\mn@eprint#1#2{\mn@eprint@#1:#2::\@nil}
\def\mn@eprint@arXiv#1{\href {http://arxiv.org/abs/#1} {{\tt arXiv:#1}}}
\def\mn@eprint@dblp#1{\href {http://dblp.uni-trier.de/rec/bibtex/#1.xml}
  {dblp:#1}}
\def\mn@eprint@#1:#2:#3:#4\@nil{\def\@tempa {#1}\def\@tempb {#2}\def\@tempc
  {#3}\ifx \@tempc \@empty \let \@tempc \@tempb \let \@tempb \@tempa \fi \ifx
  \@tempb \@empty \def\@tempb {arXiv}\fi \@ifundefined
  {mn@eprint@\@tempb}{\@tempb:\@tempc}{\expandafter \expandafter \csname
  mn@eprint@\@tempb\endcsname \expandafter{\@tempc}}}

\bibitem[\protect\citeauthoryear{{Abbott} et~al.,}{{Abbott}
  et~al.}{2017}]{abb17}
{Abbott} B.~P.,  et~al., 2017, \mn@doi [\prl] {10.1103/PhysRevLett.119.161101},
  \href {https://ui.adsabs.harvard.edu/abs/2017PhRvL.119p1101A} {119, 161101}

\bibitem[\protect\citeauthoryear{{Ablimit}, {Podsiadlowski}, {Hirai}  \&
  {Wicker}}{{Ablimit} et~al.}{2022}]{abl22}
{Ablimit} I.,  {Podsiadlowski} {\relax Ph}.,  {Hirai} R.,   {Wicker} J.,  2022,
  \mn@doi [\mnras] {10.1093/mnras/stac631}, \href
  {https://ui.adsabs.harvard.edu/abs/2022MNRAS.513.4802A} {513, 4802}

\bibitem[\protect\citeauthoryear{{Bailes}, {Lyne}  \& {Shemar}}{{Bailes}
  et~al.}{1991}]{bai91}
{Bailes} M.,  {Lyne} A.~G.,   {Shemar} S.~L.,  1991, \mn@doi [\nat]
  {10.1038/352311a0}, \href
  {https://ui.adsabs.harvard.edu/abs/1991Natur.352..311B} {352, 311}

\bibitem[\protect\citeauthoryear{{Beasor}, {Davies}, {Cabrera-Ziri}  \&
  {Hurst}}{{Beasor} et~al.}{2018}]{bea18}
{Beasor} E.~R.,  {Davies} B.,  {Cabrera-Ziri} I.,   {Hurst} G.,  2018, \mn@doi
  [\mnras] {10.1093/mnras/sty1744}, \href
  {https://ui.adsabs.harvard.edu/abs/2018MNRAS.479.3101B} {479, 3101}

\bibitem[\protect\citeauthoryear{{Biehle}}{{Biehle}}{1991}]{bie91}
{Biehle} G.~T.,  1991, \mn@doi [\apj] {10.1086/170572}, \href
  {https://ui.adsabs.harvard.edu/abs/1991ApJ...380..167B} {380, 167}

\bibitem[\protect\citeauthoryear{{Biehle}}{{Biehle}}{1994}]{bie94}
{Biehle} G.~T.,  1994, \mn@doi [\apj] {10.1086/173566}, \href
  {https://ui.adsabs.harvard.edu/abs/1994ApJ...420..364B} {420, 364}

\bibitem[\protect\citeauthoryear{{Blaauw}}{{Blaauw}}{1961}]{bla61}
{Blaauw} A.,  1961, \bain, \href
  {https://ui.adsabs.harvard.edu/abs/1961BAN....15..265B} {15, 265}

\bibitem[\protect\citeauthoryear{{Brandt} \& {Podsiadlowski}}{{Brandt} \&
  {Podsiadlowski}}{1995}]{bra95}
{Brandt} N.,  {Podsiadlowski} Ph.,  1995, \mn@doi [\mnras]
  {10.1093/mnras/274.2.461}, \href
  {https://ui.adsabs.harvard.edu/abs/1995MNRAS.274..461B} {274, 461}

\bibitem[\protect\citeauthoryear{{Bray} \& {Eldridge}}{{Bray} \&
  {Eldridge}}{2016}]{bra16}
{Bray} J.~C.,  {Eldridge} J.~J.,  2016, \mn@doi [\mnras]
  {10.1093/mnras/stw1275}, \href
  {https://ui.adsabs.harvard.edu/abs/2016MNRAS.461.3747B} {461, 3747}

\bibitem[\protect\citeauthoryear{{Bromley}, {Kenyon}, {Brown}  \&
  {Geller}}{{Bromley} et~al.}{2009}]{bro09}
{Bromley} B.~C.,  {Kenyon} S.~J.,  {Brown} W.~R.,   {Geller} M.~J.,  2009,
  \mn@doi [\apj] {10.1088/0004-637X/706/2/925}, \href
  {https://ui.adsabs.harvard.edu/abs/2009ApJ...706..925B} {706, 925}

\bibitem[\protect\citeauthoryear{{Brown}}{{Brown}}{2015}]{bro15}
{Brown} W.~R.,  2015, \mn@doi [\araa] {10.1146/annurev-astro-082214-122230},
  \href {https://ui.adsabs.harvard.edu/abs/2015ARA&A..53...15B} {53, 15}

\bibitem[\protect\citeauthoryear{{Brown}, {Geller}, {Kenyon}  \&
  {Kurtz}}{{Brown} et~al.}{2005}]{bro05}
{Brown} W.~R.,  {Geller} M.~J.,  {Kenyon} S.~J.,   {Kurtz} M.~J.,  2005,
  \mn@doi [\apjl] {10.1086/429378}, \href
  {https://ui.adsabs.harvard.edu/abs/2005ApJ...622L..33B} {622, L33}

\bibitem[\protect\citeauthoryear{{Brown}, {Geller}  \& {Kenyon}}{{Brown}
  et~al.}{2014}]{bro14}
{Brown} W.~R.,  {Geller} M.~J.,   {Kenyon} S.~J.,  2014, \mn@doi [\apj]
  {10.1088/0004-637X/787/1/89}, \href
  {https://ui.adsabs.harvard.edu/abs/2014ApJ...787...89B} {787, 89}

\bibitem[\protect\citeauthoryear{{Cannon}, {Eggleton}, {Zytkow}  \&
  {Podsiadlowski}}{{Cannon} et~al.}{1992}]{can92}
{Cannon} R.~C.,  {Eggleton} P.~P.,  {Zytkow} A.~N.,   {Podsiadlowski} {\relax
  Ph}.,  1992, \mn@doi [\apj] {10.1086/171006}, \href
  {https://ui.adsabs.harvard.edu/abs/1992ApJ...386..206C} {386, 206}

\bibitem[\protect\citeauthoryear{{Cano}}{{Cano}}{2013}]{can13}
{Cano} Z.,  2013, \mn@doi [\mnras] {10.1093/mnras/stt1048}, \href
  {https://ui.adsabs.harvard.edu/abs/2013MNRAS.434.1098C} {434, 1098}

\bibitem[\protect\citeauthoryear{{Chugai} \& {Utrobin}}{{Chugai} \&
  {Utrobin}}{2022}]{chu22}
{Chugai} N.~N.,  {Utrobin} V.~P.,  2022, \mn@doi [\mnras]
  {10.1093/mnrasl/slab131}, \href
  {https://ui.adsabs.harvard.edu/abs/2022MNRAS.512L..71C} {512, L71}

\bibitem[\protect\citeauthoryear{{De}, {MacLeod}, {Everson}, {Antoni}, {Mandel}
   \& {Ramirez-Ruiz}}{{De} et~al.}{2020}]{de20}
{De} S.,  {MacLeod} M.,  {Everson} R.~W.,  {Antoni} A.,  {Mandel} I.,
  {Ramirez-Ruiz} E.,  2020, \mn@doi [\apj] {10.3847/1538-4357/ab9ac6}, \href
  {https://ui.adsabs.harvard.edu/abs/2020ApJ...897..130D} {897, 130}

\bibitem[\protect\citeauthoryear{{DeMarchi}, {Sanders}  \&
  {Levesque}}{{DeMarchi} et~al.}{2021}]{dem21}
{DeMarchi} L.,  {Sanders} J.~R.,   {Levesque} E.~M.,  2021, \mn@doi [\apj]
  {10.3847/1538-4357/abebe1}, \href
  {https://ui.adsabs.harvard.edu/abs/2021ApJ...911..101D} {911, 101}

\bibitem[\protect\citeauthoryear{{Dexter} \& {Kasen}}{{Dexter} \&
  {Kasen}}{2013}]{dex13}
{Dexter} J.,  {Kasen} D.,  2013, \mn@doi [\apj] {10.1088/0004-637X/772/1/30},
  \href {https://ui.adsabs.harvard.edu/abs/2013ApJ...772...30D} {772, 30}

\bibitem[\protect\citeauthoryear{{Gomez}, {Berger}, {Hosseinzadeh},
  {Blanchard}, {Nicholl}  \& {Villar}}{{Gomez} et~al.}{2021}]{gom21}
{Gomez} S.,  {Berger} E.,  {Hosseinzadeh} G.,  {Blanchard} P.~K.,  {Nicholl}
  M.,   {Villar} V.~A.,  2021, \mn@doi [\apj] {10.3847/1538-4357/abf5e3}, \href
  {https://ui.adsabs.harvard.edu/abs/2021ApJ...913..143G} {913, 143}

\bibitem[\protect\citeauthoryear{{Guillochon} \& {Ramirez-Ruiz}}{{Guillochon}
  \& {Ramirez-Ruiz}}{2013}]{gui13}
{Guillochon} J.,  {Ramirez-Ruiz} E.,  2013, \mn@doi [\apj]
  {10.1088/0004-637X/767/1/25}, \href
  {https://ui.adsabs.harvard.edu/abs/2013ApJ...767...25G} {767, 25}

\bibitem[\protect\citeauthoryear{{Heber}, {Edelmann}, {Napiwotzki}, {Altmann}
  \& {Scholz}}{{Heber} et~al.}{2008}]{heb08}
{Heber} U.,  {Edelmann} H.,  {Napiwotzki} R.,  {Altmann} M.,   {Scholz} R.~D.,
  2008, \mn@doi [\aap] {10.1051/0004-6361:200809767}, \href
  {https://ui.adsabs.harvard.edu/abs/2008A&A...483L..21H} {483, L21}

\bibitem[\protect\citeauthoryear{{Hills}}{{Hills}}{1988}]{hil88}
{Hills} J.~G.,  1988, \mn@doi [\nat] {10.1038/331687a0}, \href
  {https://ui.adsabs.harvard.edu/abs/1988Natur.331..687H} {331, 687}

\bibitem[\protect\citeauthoryear{{Hirai} \& {Mandel}}{{Hirai} \&
  {Mandel}}{2021}]{RH21b}
{Hirai} R.,  {Mandel} I.,  2021, \mn@doi [\pasa] {10.1017/pasa.2021.53}, \href
  {https://ui.adsabs.harvard.edu/abs/2021PASA...38...56H} {38, e056}

\bibitem[\protect\citeauthoryear{{Hirai} \& {Yamada}}{{Hirai} \&
  {Yamada}}{2015}]{RH15}
{Hirai} R.,  {Yamada} S.,  2015, \mn@doi [\apj] {10.1088/0004-637X/805/2/170},
  \href {https://ui.adsabs.harvard.edu/abs/2015ApJ...805..170H} {805, 170}

\bibitem[\protect\citeauthoryear{{Hirai}, {Nagakura}, {Okawa}  \&
  {Fujisawa}}{{Hirai} et~al.}{2016}]{RH16}
{Hirai} R.,  {Nagakura} H.,  {Okawa} H.,   {Fujisawa} K.,  2016, \mn@doi [\prd]
  {10.1103/PhysRevD.93.083006}, \href
  {https://ui.adsabs.harvard.edu/abs/2016PhRvD..93h3006H} {93, 083006}

\bibitem[\protect\citeauthoryear{{Hirai}, {Podsiadlowski}  \& {Yamada}}{{Hirai}
  et~al.}{2018}]{RH18}
{Hirai} R.,  {Podsiadlowski} {\relax Ph}.,   {Yamada} S.,  2018, \mn@doi [\apj]
  {10.3847/1538-4357/aad6a0}, \href
  {https://ui.adsabs.harvard.edu/abs/2018ApJ...864..119H} {864, 119}

\bibitem[\protect\citeauthoryear{{Hirai}, {Podsiadlowski}, {Owocki},
  {Schneider}  \& {Smith}}{{Hirai} et~al.}{2021}]{RH21}
{Hirai} R.,  {Podsiadlowski} {\relax Ph}.,  {Owocki} S.~P.,  {Schneider} F.
  R.~N.,   {Smith} N.,  2021, \mn@doi [\mnras] {10.1093/mnras/stab571}, \href
  {https://ui.adsabs.harvard.edu/abs/2021MNRAS.503.4276H} {503, 4276}

\bibitem[\protect\citeauthoryear{{Hobbs}, {Lorimer}, {Lyne}  \&
  {Kramer}}{{Hobbs} et~al.}{2005}]{hob05}
{Hobbs} G.,  {Lorimer} D.~R.,  {Lyne} A.~G.,   {Kramer} M.,  2005, \mn@doi
  [\mnras] {10.1111/j.1365-2966.2005.09087.x}, \href
  {https://ui.adsabs.harvard.edu/abs/2005MNRAS.360..974H} {360, 974}

\bibitem[\protect\citeauthoryear{{Hober}, {Bear}  \& {Soker}}{{Hober}
  et~al.}{2022}]{hob22}
{Hober} O.,  {Bear} E.,   {Soker} N.,  2022, arXiv e-prints, \href
  {https://ui.adsabs.harvard.edu/abs/2022arXiv220511059H} {p. arXiv:2205.11059}

\bibitem[\protect\citeauthoryear{{Hosseinzadeh}, {Berger}, {Metzger}, {Gomez},
  {Nicholl}  \& {Blanchard}}{{Hosseinzadeh} et~al.}{2022}]{hos22}
{Hosseinzadeh} G.,  {Berger} E.,  {Metzger} B.~D.,  {Gomez} S.,  {Nicholl} M.,
   {Blanchard} P.,  2022, \mn@doi [\apj] {10.3847/1538-4357/ac67dd}, \href
  {https://ui.adsabs.harvard.edu/abs/2022ApJ...933...14H} {933, 14}

\bibitem[\protect\citeauthoryear{{Hoyle} \& {Lyttleton}}{{Hoyle} \&
  {Lyttleton}}{1939}]{hoy39}
{Hoyle} F.,  {Lyttleton} R.~A.,  1939, \mn@doi [Proceedings of the Cambridge
  Philosophical Society] {10.1017/S0305004100021150}, \href
  {https://ui.adsabs.harvard.edu/abs/1939PCPS...35..405H} {35, 405}

\bibitem[\protect\citeauthoryear{{Hulse} \& {Taylor}}{{Hulse} \&
  {Taylor}}{1975}]{hul75}
{Hulse} R.~A.,  {Taylor} J.~H.,  1975, \mn@doi [\apjl] {10.1086/181708}, \href
  {https://ui.adsabs.harvard.edu/abs/1975ApJ...195L..51H} {195, L51}

\bibitem[\protect\citeauthoryear{{Igoshev}, {Chruslinska}, {Dorozsmai}  \&
  {Toonen}}{{Igoshev} et~al.}{2021}]{igo21}
{Igoshev} A.~P.,  {Chruslinska} M.,  {Dorozsmai} A.,   {Toonen} S.,  2021,
  \mn@doi [\mnras] {10.1093/mnras/stab2734}, \href
  {https://ui.adsabs.harvard.edu/abs/2021MNRAS.508.3345I} {508, 3345}

\bibitem[\protect\citeauthoryear{{Irrgang}, {Przybilla}, {Heber}, {Nieva}  \&
  {Schuh}}{{Irrgang} et~al.}{2010}]{irr10}
{Irrgang} A.,  {Przybilla} N.,  {Heber} U.,  {Nieva} M.~F.,   {Schuh} S.,
  2010, \mn@doi [\apj] {10.1088/0004-637X/711/1/138}, \href
  {https://ui.adsabs.harvard.edu/abs/2010ApJ...711..138I} {711, 138}

\bibitem[\protect\citeauthoryear{{Irrgang}, {Kreuzer}, {Heber}  \&
  {Brown}}{{Irrgang} et~al.}{2018a}]{irr18b}
{Irrgang} A.,  {Kreuzer} S.,  {Heber} U.,   {Brown} W.,  2018a, \mn@doi [\aap]
  {10.1051/0004-6361/201833315}, \href
  {https://ui.adsabs.harvard.edu/abs/2018A&A...615L...5I} {615, L5}

\bibitem[\protect\citeauthoryear{{Irrgang}, {Kreuzer}  \& {Heber}}{{Irrgang}
  et~al.}{2018b}]{irr18a}
{Irrgang} A.,  {Kreuzer} S.,   {Heber} U.,  2018b, \mn@doi [\aap]
  {10.1051/0004-6361/201833874}, \href
  {https://ui.adsabs.harvard.edu/abs/2018A&A...620A..48I} {620, A48}

\bibitem[\protect\citeauthoryear{{Irrgang}, {Geier}, {Heber}, {Kupfer}  \&
  {F{\"u}rst}}{{Irrgang} et~al.}{2019}]{irr19}
{Irrgang} A.,  {Geier} S.,  {Heber} U.,  {Kupfer} T.,   {F{\"u}rst} F.,  2019,
  \mn@doi [\aap] {10.1051/0004-6361/201935429}, \href
  {https://ui.adsabs.harvard.edu/abs/2019A&A...628L...5I} {628, L5}

\bibitem[\protect\citeauthoryear{{Ivanova}, {Justham}  \& {Ricker}}{{Ivanova}
  et~al.}{2020}]{iva20}
{Ivanova} N.,  {Justham} S.,   {Ricker} P.,  2020, {Common Envelope Evolution},
  \mn@doi{10.1088/2514-3433/abb6f0.
}

\bibitem[\protect\citeauthoryear{{Janka}}{{Janka}}{2017}]{jan17}
{Janka} H.-T.,  2017, \mn@doi [\apj] {10.3847/1538-4357/aa618e}, \href
  {https://ui.adsabs.harvard.edu/abs/2017ApJ...837...84J} {837, 84}

\bibitem[\protect\citeauthoryear{{Katsuda} et~al.,}{{Katsuda}
  et~al.}{2018}]{kat18}
{Katsuda} S.,  et~al., 2018, \mn@doi [\apj] {10.3847/1538-4357/aab092}, \href
  {https://ui.adsabs.harvard.edu/abs/2018ApJ...856...18K} {856, 18}

\bibitem[\protect\citeauthoryear{{Konacki} \& {Wolszczan}}{{Konacki} \&
  {Wolszczan}}{2003}]{kon03}
{Konacki} M.,  {Wolszczan} A.,  2003, \mn@doi [\apjl] {10.1086/377093}, \href
  {https://ui.adsabs.harvard.edu/abs/2003ApJ...591L.147K} {591, L147}

\bibitem[\protect\citeauthoryear{{Kremer}, {Lombardi}, {Lu}, {Piro}  \&
  {Rasio}}{{Kremer} et~al.}{2022}]{kre22}
{Kremer} K.,  {Lombardi} James~C. J.,  {Lu} W.,  {Piro} A.~L.,   {Rasio} F.~A.,
   2022, arXiv e-prints, \href
  {https://ui.adsabs.harvard.edu/abs/2022arXiv220112368K} {p. arXiv:2201.12368}

\bibitem[\protect\citeauthoryear{{Lai}}{{Lai}}{2001}]{lai01}
{Lai} D.,  2001, in {Blaschke} D.,  {Glendenning} N.~K.,   {Sedrakian} A.,
  eds, , Vol.~578, Physics of Neutron Star Interiors.
p.~424

\bibitem[\protect\citeauthoryear{{Leonard}}{{Leonard}}{1991}]{leo91}
{Leonard} P. J.~T.,  1991, \mn@doi [\aj] {10.1086/115704}, \href
  {https://ui.adsabs.harvard.edu/abs/1991AJ....101..562L} {101, 562}

\bibitem[\protect\citeauthoryear{{Leonard}, {Hills}  \& {Dewey}}{{Leonard}
  et~al.}{1994}]{leo94}
{Leonard} P. J.~T.,  {Hills} J.~G.,   {Dewey} R.~J.,  1994, \mn@doi [\apjl]
  {10.1086/187225}, \href
  {https://ui.adsabs.harvard.edu/abs/1994ApJ...423L..19L} {423, L19}

\bibitem[\protect\citeauthoryear{{Levesque}, {Massey}, {Zytkow}  \&
  {Morrell}}{{Levesque} et~al.}{2014}]{lev14}
{Levesque} E.~M.,  {Massey} P.,  {Zytkow} A.~N.,   {Morrell} N.,  2014, \mn@doi
  [\mnras] {10.1093/mnrasl/slu080}, \href
  {https://ui.adsabs.harvard.edu/abs/2014MNRAS.443L..94L} {443, L94}

\bibitem[\protect\citeauthoryear{{Li} et~al.,}{{Li} et~al.}{2011}]{li11}
{Li} W.,  et~al., 2011, \mn@doi [\mnras] {10.1111/j.1365-2966.2011.18160.x},
  \href {https://ui.adsabs.harvard.edu/abs/2011MNRAS.412.1441L} {412, 1441}

\bibitem[\protect\citeauthoryear{{Lyman}, {Bersier}, {James}, {Mazzali},
  {Eldridge}, {Fraser}  \& {Pian}}{{Lyman} et~al.}{2016}]{lym16}
{Lyman} J.~D.,  {Bersier} D.,  {James} P.~A.,  {Mazzali} P.~A.,  {Eldridge}
  J.~J.,  {Fraser} M.,   {Pian} E.,  2016, \mn@doi [\mnras]
  {10.1093/mnras/stv2983}, \href
  {https://ui.adsabs.harvard.edu/abs/2016MNRAS.457..328L} {457, 328}

\bibitem[\protect\citeauthoryear{{Lyne} \& {Lorimer}}{{Lyne} \&
  {Lorimer}}{1994}]{lyn94}
{Lyne} A.~G.,  {Lorimer} D.~R.,  1994, \mn@doi [\nat] {10.1038/369127a0}, \href
  {https://ui.adsabs.harvard.edu/abs/1994Natur.369..127L} {369, 127}

\bibitem[\protect\citeauthoryear{{MacLeod} \& {Ramirez-Ruiz}}{{MacLeod} \&
  {Ramirez-Ruiz}}{2015}]{mac15}
{MacLeod} M.,  {Ramirez-Ruiz} E.,  2015, \mn@doi [\apj]
  {10.1088/0004-637X/803/1/41}, \href
  {https://ui.adsabs.harvard.edu/abs/2015ApJ...803...41M} {803, 41}

\bibitem[\protect\citeauthoryear{{MacLeod}, {Antoni}, {Murguia-Berthier},
  {Macias}  \& {Ramirez-Ruiz}}{{MacLeod} et~al.}{2017}]{mac17}
{MacLeod} M.,  {Antoni} A.,  {Murguia-Berthier} A.,  {Macias} P.,
  {Ramirez-Ruiz} E.,  2017, \mn@doi [\apj] {10.3847/1538-4357/aa6117}, \href
  {https://ui.adsabs.harvard.edu/abs/2017ApJ...838...56M} {838, 56}

\bibitem[\protect\citeauthoryear{{MacLeod}, {Vick}  \& {Loeb}}{{MacLeod}
  et~al.}{2022}]{mac22}
{MacLeod} M.,  {Vick} M.,   {Loeb} A.,  2022, arXiv e-prints, \href
  {https://ui.adsabs.harvard.edu/abs/2022arXiv220301947M} {p. arXiv:2203.01947}

\bibitem[\protect\citeauthoryear{{Mayer}, {Becker}, {Patnaude}, {Winkler}  \&
  {Kraft}}{{Mayer} et~al.}{2020}]{may20}
{Mayer} M.,  {Becker} W.,  {Patnaude} D.,  {Winkler} P.~F.,   {Kraft} R.,
  2020, \mn@doi [\apj] {10.3847/1538-4357/aba121}, \href
  {https://ui.adsabs.harvard.edu/abs/2020ApJ...899..138M} {899, 138}

\bibitem[\protect\citeauthoryear{{Moriya}}{{Moriya}}{2018}]{mor18}
{Moriya} T.~J.,  2018, \mn@doi [\mnras] {10.1093/mnrasl/sly005}, \href
  {https://ui.adsabs.harvard.edu/abs/2018MNRAS.475L..49M} {475, L49}

\bibitem[\protect\citeauthoryear{{Moriya} \& {Blinnikov}}{{Moriya} \&
  {Blinnikov}}{2021}]{mor21}
{Moriya} T.~J.,  {Blinnikov} S.~I.,  2021, \mn@doi [\mnras]
  {10.1093/mnras/stab2584}, \href
  {https://ui.adsabs.harvard.edu/abs/2021MNRAS.508...74M} {508, 74}

\bibitem[\protect\citeauthoryear{{Moriya}, {Murase}, {Kashiyama}  \&
  {Blinnikov}}{{Moriya} et~al.}{2022}]{mor22}
{Moriya} T.~J.,  {Murase} K.,  {Kashiyama} K.,   {Blinnikov} S.~I.,  2022,
  \mn@doi [\mnras] {10.1093/mnras/stac1352}, \href
  {https://ui.adsabs.harvard.edu/abs/2022MNRAS.513.6210M} {513, 6210}

\bibitem[\protect\citeauthoryear{{Nakamura} \& {Piran}}{{Nakamura} \&
  {Piran}}{1991}]{nak91}
{Nakamura} T.,  {Piran} T.,  1991, \mn@doi [\apjl] {10.1086/186217}, \href
  {https://ui.adsabs.harvard.edu/abs/1991ApJ...382L..81N} {382, L81}

\bibitem[\protect\citeauthoryear{{Narayan} \& {Yi}}{{Narayan} \&
  {Yi}}{1994}]{nar94}
{Narayan} R.,  {Yi} I.,  1994, \mn@doi [\apjl] {10.1086/187381}, \href
  {https://ui.adsabs.harvard.edu/abs/1994ApJ...428L..13N} {428, L13}

\bibitem[\protect\citeauthoryear{{Nicholl} et~al.,}{{Nicholl}
  et~al.}{2016}]{nic16}
{Nicholl} M.,  et~al., 2016, \mn@doi [\apjl] {10.3847/2041-8205/828/2/L18},
  \href {https://ui.adsabs.harvard.edu/abs/2016ApJ...828L..18N} {828, L18}

\bibitem[\protect\citeauthoryear{{Ni{\c{t}}u}, {Keith}, {Stappers}, {Lyne}  \&
  {Mickaliger}}{{Ni{\c{t}}u} et~al.}{2022}]{nit22}
{Ni{\c{t}}u} I.~C.,  {Keith} M.~J.,  {Stappers} B.~W.,  {Lyne} A.~G.,
  {Mickaliger} M.~B.,  2022, \mn@doi [\mnras] {10.1093/mnras/stac593}, \href
  {https://ui.adsabs.harvard.edu/abs/2022MNRAS.512.2446N} {512, 2446}

\bibitem[\protect\citeauthoryear{{Ogata}, {Hirai}  \& {Hijikawa}}{{Ogata}
  et~al.}{2021}]{oga21}
{Ogata} M.,  {Hirai} R.,   {Hijikawa} K.,  2021, \mn@doi [\mnras]
  {10.1093/mnras/stab1439}, \href
  {https://ui.adsabs.harvard.edu/abs/2021MNRAS.505.2485O} {505, 2485}

\bibitem[\protect\citeauthoryear{{Ostriker}}{{Ostriker}}{1999}]{ost99}
{Ostriker} E.~C.,  1999, \mn@doi [\apj] {10.1086/306858}, \href
  {https://ui.adsabs.harvard.edu/abs/1999ApJ...513..252O} {513, 252}

\bibitem[\protect\citeauthoryear{{Ostriker} \& {Gunn}}{{Ostriker} \&
  {Gunn}}{1971}]{ost71}
{Ostriker} J.~P.,  {Gunn} J.~E.,  1971, \mn@doi [\apjl] {10.1086/180699}, \href
  {https://ui.adsabs.harvard.edu/abs/1971ApJ...164L..95O} {164, L95}

\bibitem[\protect\citeauthoryear{{Paxton}, {Bildsten}, {Dotter}, {Herwig},
  {Lesaffre}  \& {Timmes}}{{Paxton} et~al.}{2011}]{MESA1}
{Paxton} B.,  {Bildsten} L.,  {Dotter} A.,  {Herwig} F.,  {Lesaffre} P.,
  {Timmes} F.,  2011, \mn@doi [\apjs] {10.1088/0067-0049/192/1/3}, \href
  {https://ui.adsabs.harvard.edu/abs/2011ApJS..192....3P} {192, 3}

\bibitem[\protect\citeauthoryear{{Paxton} et~al.,}{{Paxton}
  et~al.}{2013}]{MESA2}
{Paxton} B.,  et~al., 2013, \mn@doi [\apjs] {10.1088/0067-0049/208/1/4}, \href
  {https://ui.adsabs.harvard.edu/abs/2013ApJS..208....4P} {208, 4}

\bibitem[\protect\citeauthoryear{{Paxton} et~al.,}{{Paxton}
  et~al.}{2015}]{MESA3}
{Paxton} B.,  et~al., 2015, \mn@doi [\apjs] {10.1088/0067-0049/220/1/15}, \href
  {https://ui.adsabs.harvard.edu/abs/2015ApJS..220...15P} {220, 15}

\bibitem[\protect\citeauthoryear{{Paxton} et~al.,}{{Paxton}
  et~al.}{2018}]{MESA4}
{Paxton} B.,  et~al., 2018, \mn@doi [\apjs] {10.3847/1538-4365/aaa5a8}, \href
  {https://ui.adsabs.harvard.edu/abs/2018ApJS..234...34P} {234, 34}

\bibitem[\protect\citeauthoryear{{Paxton} et~al.,}{{Paxton}
  et~al.}{2019}]{MESA5}
{Paxton} B.,  et~al., 2019, \mn@doi [\apjs] {10.3847/1538-4365/ab2241}, \href
  {https://ui.adsabs.harvard.edu/abs/2019ApJS..243...10P} {243, 10}

\bibitem[\protect\citeauthoryear{{Perets} \& {{\v{S}}ubr}}{{Perets} \&
  {{\v{S}}ubr}}{2012}]{per12}
{Perets} H.~B.,  {{\v{S}}ubr} L.,  2012, \mn@doi [\apj]
  {10.1088/0004-637X/751/2/133}, \href
  {https://ui.adsabs.harvard.edu/abs/2012ApJ...751..133P} {751, 133}

\bibitem[\protect\citeauthoryear{{Phinney} \& {Hansen}}{{Phinney} \&
  {Hansen}}{1993}]{phi93}
{Phinney} E.~S.,  {Hansen} B.~M.~S.,  1993, in {Phillips} J.~A.,  {Thorsett}
  S.~E.,   {Kulkarni} S.~R.,  eds,  Astronomical Society of the Pacific
  Conference Series Vol. 36, Planets Around Pulsars. pp 371--390

\bibitem[\protect\citeauthoryear{{Podsiadlowski}}{{Podsiadlowski}}{1993}]{pod93}
{Podsiadlowski} {\relax Ph}.,  1993, in {Phillips} J.~A.,  {Thorsett} S.~E.,
  {Kulkarni} S.~R.,  eds,  Astronomical Society of the Pacific Conference
  Series Vol. 36, Planets Around Pulsars. pp 149--165

\bibitem[\protect\citeauthoryear{{Podsiadlowski}, {Cannon}  \&
  {Rees}}{{Podsiadlowski} et~al.}{1995}]{pod95a}
{Podsiadlowski} {\relax Ph}.,  {Cannon} R.~C.,   {Rees} M.~J.,  1995, \mn@doi
  [\mnras] {10.1093/mnras/274.2.485}, \href
  {https://ui.adsabs.harvard.edu/abs/1995MNRAS.274..485P} {274, 485}

\bibitem[\protect\citeauthoryear{{Poveda}, {Ruiz}  \& {Allen}}{{Poveda}
  et~al.}{1967}]{pov67}
{Poveda} A.,  {Ruiz} J.,   {Allen} C.,  1967, Boletin de los Observatorios
  Tonantzintla y Tacubaya, \href
  {https://ui.adsabs.harvard.edu/abs/1967BOTT....4...86P} {4, 86}

\bibitem[\protect\citeauthoryear{{Price} \& {Monaghan}}{{Price} \&
  {Monaghan}}{2007}]{pri07}
{Price} D.~J.,  {Monaghan} J.~J.,  2007, \mn@doi [\mnras]
  {10.1111/j.1365-2966.2006.11241.x}, \href
  {https://ui.adsabs.harvard.edu/abs/2007MNRAS.374.1347P} {374, 1347}

\bibitem[\protect\citeauthoryear{{Quimby} et~al.,}{{Quimby}
  et~al.}{2011}]{qui11}
{Quimby} R.~M.,  et~al., 2011, \mn@doi [\nat] {10.1038/nature10095}, \href
  {https://ui.adsabs.harvard.edu/abs/2011Natur.474..487Q} {474, 487}

\bibitem[\protect\citeauthoryear{{Ray}, {Kembhavi}  \& {Antia}}{{Ray}
  et~al.}{1987}]{ray87}
{Ray} A.,  {Kembhavi} A.~K.,   {Antia} H.~M.,  1987, \aap, \href
  {https://ui.adsabs.harvard.edu/abs/1987A&A...184..164R} {184, 164}

\bibitem[\protect\citeauthoryear{{Renzo} et~al.,}{{Renzo} et~al.}{2019}]{ren19}
{Renzo} M.,  et~al., 2019, \mn@doi [\aap] {10.1051/0004-6361/201833297}, \href
  {https://ui.adsabs.harvard.edu/abs/2019A&A...624A..66R} {624, A66}

\bibitem[\protect\citeauthoryear{{Rose}, {Naoz}, {Sari}  \& {Linial}}{{Rose}
  et~al.}{2022}]{ros22}
{Rose} S.~C.,  {Naoz} S.,  {Sari} R.,   {Linial} I.,  2022, \mn@doi [\apjl]
  {10.3847/2041-8213/ac6426}, \href
  {https://ui.adsabs.harvard.edu/abs/2022ApJ...929L..22R} {929, L22}

\bibitem[\protect\citeauthoryear{{Ruffert}}{{Ruffert}}{1992}]{ruf92}
{Ruffert} M.,  1992, \aap, \href
  {https://ui.adsabs.harvard.edu/abs/1992A&A...265...82R} {265, 82}

\bibitem[\protect\citeauthoryear{{Ruffert}}{{Ruffert}}{1993}]{ruf93}
{Ruffert} M.,  1993, \aap, \href
  {https://ui.adsabs.harvard.edu/abs/1993A&A...280..141R} {280, 141}

\bibitem[\protect\citeauthoryear{{Ruffert} \& {M\"uller}}{{Ruffert} \&
  {M\"uller}}{1990}]{ruf90}
{Ruffert} M.,  {M\"uller} E.,  1990, \aap, \href
  {https://ui.adsabs.harvard.edu/abs/1990A&A...238..116R} {238, 116}

\bibitem[\protect\citeauthoryear{{Ryu}, {Krolik}, {Piran}  \& {Noble}}{{Ryu}
  et~al.}{2020}]{ryu20}
{Ryu} T.,  {Krolik} J.,  {Piran} T.,   {Noble} S.~C.,  2020, \mn@doi [\apj]
  {10.3847/1538-4357/abb3ce}, \href
  {https://ui.adsabs.harvard.edu/abs/2020ApJ...904..100R} {904, 100}

\bibitem[\protect\citeauthoryear{{Sana} et~al.,}{{Sana} et~al.}{2012}]{san12}
{Sana} H.,  et~al., 2012, \mn@doi [Science] {10.1126/science.1223344}, \href
  {https://ui.adsabs.harvard.edu/abs/2012Sci...337..444S} {337, 444}

\bibitem[\protect\citeauthoryear{{Sana} et~al.,}{{Sana} et~al.}{2014}]{san14}
{Sana} H.,  et~al., 2014, \mn@doi [\apjs] {10.1088/0067-0049/215/1/15}, \href
  {https://ui.adsabs.harvard.edu/abs/2014ApJS..215...15S} {215, 15}

\bibitem[\protect\citeauthoryear{{Smith}, {Li}, {Filippenko}  \&
  {Chornock}}{{Smith} et~al.}{2011}]{smi11}
{Smith} N.,  {Li} W.,  {Filippenko} A.~V.,   {Chornock} R.,  2011, \mn@doi
  [\mnras] {10.1111/j.1365-2966.2011.17229.x}, \href
  {https://ui.adsabs.harvard.edu/abs/2011MNRAS.412.1522S} {412, 1522}

\bibitem[\protect\citeauthoryear{{Soker}}{{Soker}}{2016}]{sok16}
{Soker} N.,  2016, \mn@doi [\na] {10.1016/j.newast.2016.02.009}, \href
  {https://ui.adsabs.harvard.edu/abs/2016NewA...47...88S} {47, 88}

\bibitem[\protect\citeauthoryear{{Soker}}{{Soker}}{2022}]{sok22}
{Soker} N.,  2022, arXiv e-prints, \href
  {https://ui.adsabs.harvard.edu/abs/2022arXiv220509560S} {p. arXiv:2205.09560}

\bibitem[\protect\citeauthoryear{{Sorokina}, {Blinnikov}, {Nomoto}, {Quimby}
  \& {Tolstov}}{{Sorokina} et~al.}{2016}]{sor16}
{Sorokina} E.,  {Blinnikov} S.,  {Nomoto} K.,  {Quimby} R.,   {Tolstov} A.,
  2016, \mn@doi [\apj] {10.3847/0004-637X/829/1/17}, \href
  {https://ui.adsabs.harvard.edu/abs/2016ApJ...829...17S} {829, 17}

\bibitem[\protect\citeauthoryear{{Taam}, {Bodenheimer}  \& {Ostriker}}{{Taam}
  et~al.}{1978}]{taa78}
{Taam} R.~E.,  {Bodenheimer} P.,   {Ostriker} J.~P.,  1978, \mn@doi [\apj]
  {10.1086/156142}, \href
  {https://ui.adsabs.harvard.edu/abs/1978ApJ...222..269T} {222, 269}

\bibitem[\protect\citeauthoryear{{Taddia} et~al.,}{{Taddia}
  et~al.}{2015}]{tad15}
{Taddia} F.,  et~al., 2015, \mn@doi [\aap] {10.1051/0004-6361/201423915}, \href
  {https://ui.adsabs.harvard.edu/abs/2015A&A...574A..60T} {574, A60}

\bibitem[\protect\citeauthoryear{{Taddia} et~al.,}{{Taddia}
  et~al.}{2018}]{tad18}
{Taddia} F.,  et~al., 2018, \mn@doi [\aap] {10.1051/0004-6361/201730844}, \href
  {https://ui.adsabs.harvard.edu/abs/2018A&A...609A.136T} {609, A136}

\bibitem[\protect\citeauthoryear{{Tauris}}{{Tauris}}{2015}]{tau15}
{Tauris} T.~M.,  2015, \mn@doi [\mnras] {10.1093/mnrasl/slu189}, \href
  {https://ui.adsabs.harvard.edu/abs/2015MNRAS.448L...6T} {448, L6}

\bibitem[\protect\citeauthoryear{{Tauris} \& {Takens}}{{Tauris} \&
  {Takens}}{1998}]{tau98}
{Tauris} T.~M.,  {Takens} R.~J.,  1998, \aap, \href
  {https://ui.adsabs.harvard.edu/abs/1998A&A...330.1047T} {330, 1047}

\bibitem[\protect\citeauthoryear{{Tauris} \& {van den Heuvel}}{{Tauris} \& {van
  den Heuvel}}{2006}]{tau06}
{Tauris} T.~M.,  {van den Heuvel} E.~P.~J.,  2006, in , Vol.~39, Compact
  stellar X-ray sources.
pp 623--665

\bibitem[\protect\citeauthoryear{{Thorne} \& {\.Zytkow}}{{Thorne} \&
  {\.Zytkow}}{1975}]{tho75}
{Thorne} K.~S.,  {\.Zytkow} A.~N.,  1975, \mn@doi [\apjl] {10.1086/181839},
  \href {https://ui.adsabs.harvard.edu/abs/1975ApJ...199L..19T} {199, L19}

\bibitem[\protect\citeauthoryear{{Thorne} \& {\.Zytkow}}{{Thorne} \&
  {\.Zytkow}}{1977}]{tho77}
{Thorne} K.~S.,  {\.Zytkow} A.~N.,  1977, \mn@doi [\apj] {10.1086/155109},
  \href {https://ui.adsabs.harvard.edu/abs/1977ApJ...212..832T} {212, 832}

\bibitem[\protect\citeauthoryear{{Verbunt}, {Igoshev}  \& {Cator}}{{Verbunt}
  et~al.}{2017}]{ver17}
{Verbunt} F.,  {Igoshev} A.,   {Cator} E.,  2017, \mn@doi [\aap]
  {10.1051/0004-6361/201731518}, \href
  {https://ui.adsabs.harvard.edu/abs/2017A&A...608A..57V} {608, A57}

\bibitem[\protect\citeauthoryear{{Wang}, {Chakrabarty}  \& {Kaplan}}{{Wang}
  et~al.}{2006}]{wan06}
{Wang} Z.,  {Chakrabarty} D.,   {Kaplan} D.~L.,  2006, \mn@doi [\nat]
  {10.1038/nature04669}, \href
  {https://ui.adsabs.harvard.edu/abs/2006Natur.440..772W} {440, 772}

\bibitem[\protect\citeauthoryear{{Wolszczan} \& {Frail}}{{Wolszczan} \&
  {Frail}}{1992}]{wol92}
{Wolszczan} A.,  {Frail} D.~A.,  1992, \mn@doi [\nat] {10.1038/355145a0}, \href
  {https://ui.adsabs.harvard.edu/abs/1992Natur.355..145W} {355, 145}

\bibitem[\protect\citeauthoryear{{Zapartas} et~al.,}{{Zapartas}
  et~al.}{2017}]{zap17}
{Zapartas} E.,  et~al., 2017, \mn@doi [\apj] {10.3847/1538-4357/aa7467}, \href
  {https://ui.adsabs.harvard.edu/abs/2017ApJ...842..125Z} {842, 125}

\makeatother
\end{thebibliography}

%%%%%%%%%%%%%%%%%%%%%%%%%%%%%%%%%%%%%%%%%%%%%%%%%%

%%%%%%%%%%%%%%%%% APPENDICES %%%%%%%%%%%%%%%%%%%%%

\appendix

%\section{Analytical model for drag in a star}\label{app:drag_model}

%%%%%%%%%%%%%%%%%%%%%%%%%%%%%%%%%%%%%%%%%%%%%%%%%%

% Don't change these lines
\bsp	% typesetting comment
\label{lastpage}
\end{document}